\documentclass[reprint,
superscriptaddress,
nofootinbib,
 amsmath,amssymb,
 aps,
pra,
a4paper,twocolumn, floatfix
]{revtex4-2}

\usepackage{dsfont}
\usepackage{graphicx}%
\usepackage{dcolumn}%
\usepackage{bm}%
\usepackage[hidelinks]{hyperref}%

\hypersetup{
    colorlinks,
    linkcolor={red!50!black},
    citecolor={blue!50!black},
    urlcolor={blue!80!black}
}
\usepackage{comment}
\newcommand{\ket}[1]{|#1\rangle}

\usepackage{physics}
\usepackage{amsmath}
\usepackage{amsthm}
\usepackage{amssymb}
\usepackage{nicefrac} 
\usepackage{tikz}
\usetikzlibrary{matrix,backgrounds,quantikz2}

\usepackage{booktabs}
\usepackage{multirow}

\usepackage{mathtools}
\mathtoolsset{showonlyrefs=true}

\begin{document}

\begin{abstract}

Knowing whether a Quantum Machine Learning model would perform well on a given dataset before training it can help to save critical resources. 
However, gathering a priori information about model performance (e.g., training speed, critical hyperparameters, or inference capabilities on unseen data) is a highly non-trivial task, in general. 
Recently, the Quantum Neural Tangent Kernel (QNTK) has been proposed as a powerful mathematical tool to describe the behavior of Quantum Neural Network (QNN) models.
In this work, we propose a practical framework allowing to employ the QNTK for QNN performance diagnostics. More specifically, we show how a critical learning rate and a characteristic decay time for the average training error can be estimated from the spectrum of the QNTK evaluated at the initialization stage. We then show how a QNTK-based kernel formula can be used to analyze, up to a first-order approximation, the expected inference capabilities of the quantum model under study. 
We validate our proposed approach with extensive numerical simulations, using different QNN architectures and datasets.
Our results demonstrate that QNTK diagnostics yields accurate approximations of QNN behavior for sufficiently deep circuits, can provide insights for shallow QNNs, and enables detecting -- hence also addressing -- potential shortcomings in model design. 
\end{abstract}

\title{Towards Practical Quantum Neural Network Diagnostics \\ with Neural Tangent Kernels}

\author{Francesco Scala}
\email{francesco.scala01@ateneopv.it}
 \affiliation{Dipartimento di Fisica,  Università di Pavia, via Bassi 6, 27100 Pavia, Italy}

\author{Christa Zoufal}
 \affiliation{IBM Quantum, IBM Research--Zurich, 8803 Rüschlikon, Switzerland}

\author{Dario Gerace}
 \affiliation{Dipartimento di Fisica,  Università di Pavia, via Bassi 6, 27100 Pavia, Italy}

\author{Francesco Tacchino}
\email{fta@zurich.ibm.com}
 \affiliation{IBM Quantum, IBM Research--Zurich, 8803 Rüschlikon, Switzerland}

\date{\today}

\maketitle

\section{Introduction}
\label{sec:intro}

In recent years, rapid advances in Machine Learning (ML)~\cite{tyson2020MachineLearning,amanpreetReviewSupervised2016,shinde2018review,Dhall2019ML,ray2023gpt,Jumper2021alphafold} and quantum computing technologies~\cite{Tacchino2020AQT,bravyi2022future,kim2023evidence} happened in parallel, thus opening novel research avenues and bringing up a variety of new questions. 
The field that investigates potential synergies of ML and quantum computing is generally referred to as  \textit{Quantum Machine Learning} (QML)~\cite{2014QMLWittek,Biamonte2017quantum,cerezo2022challenges}. 
One the most important questions within this field is whether quantum information processing techniques can be leveraged to solve certain learning problems more effectively than performing the same tasks by solely using classical resources~\cite{LiuRigorousRobustQML21, gyurik2023exponentialseparationsclassicalquantum,Huang2021power,abbas2021power,sweke2023potential,bowles2024betterclassicalsubtleart}.  
In this context, various QML algorithms based on Parametrized Quantum Circuits~(PQCs)~\cite{Benedetti2019pqc,Peruzzo2014VQE,tacchino2021variational,Cerezo2021VQA}, such as Quantum Neural Networks~(QNNs)~\cite{Mangini2021qnns,abbas2021power} and quantum kernel methods~\cite{Schuld2019_kernel,Havlicek2019_nature,glick2024covariant}, were proposed and investigated. The properties of these models, including expressivity~\cite{schuld2021effect}, entangling power~\cite{Sim2019,Hubregtsen2021,Ballarin_2023}, generalization capabilities~\cite{caro2022generalization,peters2023generalization,gilfuster2024understanding}, overparametrization~\cite{haug2021capacity,larocca2023theory}, and trainability~\cite{McClean2018barren,holmes2022connecting,Arrasmith2022equivalence,larocca2024reviewBPs} have been extensively studied.
Nevertheless, it is still a non-trivial challenge to determine how a specific quantum model will perform on a given dataset before the actual training is executed and the model applied.
Notably, model performance is related to training speed~\cite{abbas2021power}, ability to yield correct predictions on previously unseen data~\cite{caro2022generalization, gilfuster2024understanding}, avoidance of overfitting~\cite{hawkins2004problem, ying2019overview}, and necessity of regularization procedures~\cite{tian2022regularization,elemstatlear,wan2013dropconnect,srivastava2014dropout,kobayashi2022entangling_drop,scala2023dropout}.

In classical ML, many of the above questions have been investigated by developing effective theories for deep learning~\cite{roberts2022principles} and by introducing the concept of the Neural Tangent Kernel (NTK)~\cite{jacot2018ntk,lee2019wide,arora2019exact,xiao2020disentangling, karakida2020pathological,yang2020spectral,adlam2020tripledescent}. %
Inspired by this idea, evaluating the NTK for quantum models, thereby constructing a so-called \textit{Quantum Neural Tangent Kernel} (QNTK), was recently proposed as a mathematical tool to investigate the training behavior of QNNs~\cite{liu2022representation,liu2023analytic, Abedi2023quantumlazytraining, wang2023symmetric,zhang2024dynamical,shirai2022quantum,incudini2023path,wang2024comprehensive}. This allowed, for instance, to prove analytically that the residual training errors decay exponentially for deep random~\cite{liu2022representation, liu2023analytic} or symmetric~\cite{wang2023symmetric} QNNs. Further investigations uncovered the details of single-trajectory dynamics beyond average cases~\cite{you2023analyzing}, clarified how similar mechanisms apply to the Variational Quantum Eigensolver (VQE) algorithm under Riemannian Gradient Flow~\cite{you2022convergence} and discussed QNTK concentration effects with respect to the distribution of the inputs~\cite{yu2024expressibility}.
However, despite the many theoretical studies, only little attention has been devoted to the practical use of the QNTK as a diagnostic tool for QML models or as an instrument to guide model design, so far.
In this work, we address such open questions through a combination of analytic and numerical techniques.  

More specifically, we demonstrate how the QNTK can be used before any training takes place (i.e., at initialization) to gain a qualitative understanding of the expected performance of a QML algorithm on a concrete problem instance. 
In analogy to the classical literature~\cite{xiao2020disentangling}, we identify the condition number of the QNTK matrix as the relevant quantity to estimate the characteristic decay time of the average training error. 
Furthermore, we connect the maximum eigenvalue of the QNTK to a critical learning rate, which ensures a stable and smooth training convergence~\cite{lewkowycz2020catapult}. 
Additionally, we show how the QNTK can be leveraged to approximate expected QNN predictions on test data, which in turn allows us to gauge the inference capabilities of the model. %
We validate our theoretical results with extensive numerical simulations on multiple QNN architectures and different datasets. Importantly, most of the formal results are obtained under specific assumptions, i.e., the so called \textit{lazy training} regime, which are typically difficult to realize in practice. Interestingly, however, we  empirically find that many of our predictions also approximately hold under more realistic conditions, e.g., for certain shallow models.

The manuscript is organized as follows: in Sec.~\ref{sec:background}, we briefly recall the fundamental concepts and results from QNTK theory; in Sec.~\ref{sec:diagnostic}, we derive the main analytical results of this work, and show how to use the QNTK at initialization as a QNN diagnostic tool; in Sec.~\ref{sec:results} we describe the results of our numerical simulations; finally, we draw conclusions in Sec.~\ref{sec:conclusions}.

\begin{figure*}
    \centering
    \includegraphics[trim={3.cm 5.cm 3.cm 5.cm},clip,width=.8\linewidth]{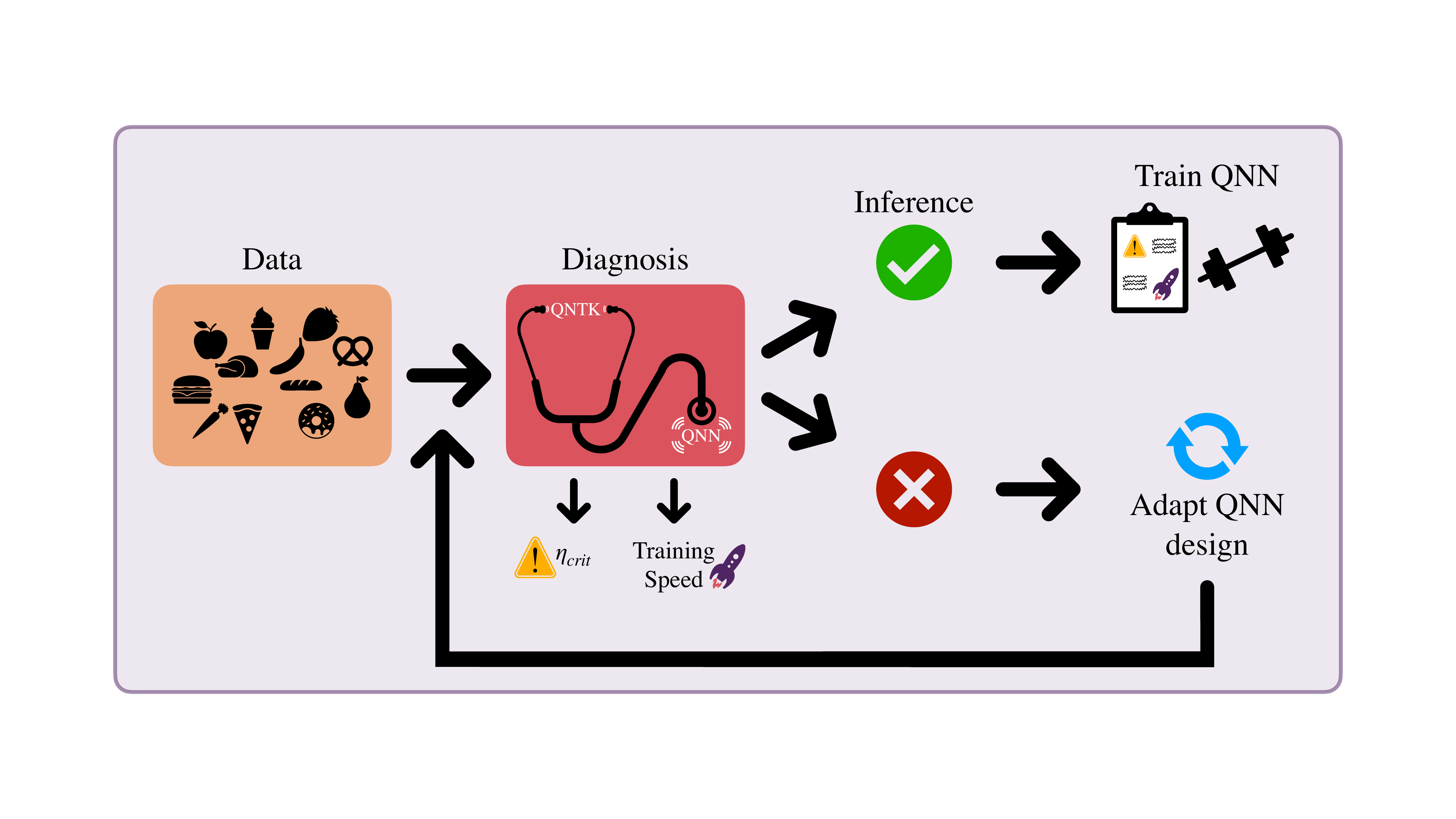}
    \caption{%
    Schematic representation of the diagnostic procedure of a Quantum Neural Network (QNN) model by means of the Quantum Neural Tangent Kernel (QNTK). By evaluating the latter on a specific training set at the model initialization stage, it is possible to collect useful information on the training dynamics (critical learning rate and training speed) and to assess the QNN inference capabilities. These can then be used to set the correct training hyperparameters, or to detect potential shortcomings in the model design.}
    \label{fig:qntk_scheme}
\end{figure*}

\section{Theoretical Background}
\label{sec:background}
In this section, we introduce the fundamental concepts of QNTK theory in a supervised learning setting. More detailed derivations can be found in App.~\ref{appendix:NTK} (for completeness) and Refs.~\cite{jacot2018ntk, liu2022representation, liu2023analytic, shirai2022quantum, roberts2022principles}.
\\
Given a dataset $\left\{\mathbf{x}_i, y_i\right\}_{i=1}^{M}$ with points $\left\{\mathbf{x}_i\right\}_{i=1}^{M}$ sampled from a data distribution $\mathcal{D}$ and corresponding labels $\left\{y_i\right\}_{i=1}^{M}$, we define a QNN model by fixing an observable (i.e., a Hermitian operator)  $O$ and evaluating its expectation value with respect to a parameterized quantum state $\ket{\psi(\mathbf{x}_i, \boldsymbol{\theta})}$ as:
\begin{equation}
    f(\mathbf{x}_i, \boldsymbol{\theta})=\langle\psi(\mathbf{x}_i, \boldsymbol{\theta})| O \ket{\psi(\mathbf{x}_i, \boldsymbol{\theta})}, 
\end{equation}
with
\begin{equation}
    \ket{\psi(\mathbf{x}_i, \boldsymbol{\theta})} = U(\mathbf{x}_i, \boldsymbol{\theta})\ket{0}=\prod_{l=0}^L U_l(\boldsymbol{\theta})S_l(\mathbf{x}_i)\ket{0}.
\end{equation}
Here, $U_l(\boldsymbol{\theta})$ are parametrized trainable unitaries, $S_l(\mathbf{x}_i)$ represent encoding operations that embed the classical data points $\mathbf{x}_i$, and $L$ is the number of layers (i.e., the model depth) present in the QNN. Notice that, even though for simplicity we will only consider scalar output functions, this model could be easily generalised to output a vector $\boldsymbol{f}$ by assigning a different operator $O_j$ to each desired output component $f_j$~\cite{liu2022representation, liu2023analytic}. As the number of layers $L$ is increased, thereby introducing more and more independent trainable parameters, one can eventually reach an \textit{overparametrized} regime for the QNN model. Informally, this happens when the number of independent parameters exceeds the number of degrees of freedom of the Hilbert (sub-)space accessible to the model ansatz. %
Formal definitions of overparametrization can be given w.r.t.~the quantum Fisher information matrix rank~\cite{haug2021capacity} and the dimension of the dynamical Lie algebra~\cite{larocca2023theory}. 

To derive an expression for the QNTK, let us assume that the following quadratic loss function (also known as the Mean Squared Error, MSE) is used to train our QNN model:
\begin{equation}
\label{eq:loss_ntk}
    \mathcal{L(\boldsymbol{\theta})}= \frac{1}{2}\sum _{i=1}^M(f(\mathbf{x}_i,\boldsymbol{\theta})-y_{i})^2=\frac{1}{2}\sum _{i}\varepsilon_{i}^2.
\end{equation}
During training, the parameters $\boldsymbol{\theta}$ are iteratively adjusted in order to minimize $\mathcal{L}$. To this purpose, a standard gradient descent algorithm can be applied, such that the $l$-th variational parameter is updated at the $(t+1)$-th step as
\begin{align}
    \delta\theta_l  &= \theta_l(t+1)-\theta_l(t)=-\eta \frac{\partial \mathcal{L}}{\partial\theta_l}= -\eta \sum _{\alpha'}\varepsilon_{\alpha'}\frac{\partial \varepsilon_{\alpha'}}{\partial\theta_l}.
\end{align}
Here, $\eta$ defines the \textit{learning rate} which sets the overall scale of the updates. If we choose $\eta$ to be sufficiently small, we can approximate the corresponding change in the training error as
\begin{equation}
\label{eq:training error update}
    \delta \varepsilon_{i} =  
    \varepsilon_{i}(t+1)-\varepsilon_{i}(t)=
    -\eta \sum _{i'} K_{ii'}\varepsilon_{i'},
\end{equation}
in which
\begin{equation}    
\label{eq:QNTK}
    K_{ii'}=\sum_l \frac{\partial \varepsilon_{i}}{\partial\theta_l} \frac{\partial \varepsilon_{i'}}{\partial\theta_l},
\end{equation}
is the QNTK associated with the given quantum model, $f$. Notice that $K$ is a symmetric, positive semi-definite matrix of dimensions $ M\times M$ that, interestingly, has also been proposed as a trainable quantum kernel to be employed in quantum support vector machines~\cite{shirai2022quantum, incudini2023path}.

Similarly to its classical equivalent, the QNTK is a particularly convenient object to characterize the training dynamics of a model whenever the latter can be approximated by its linearization in the vicinity of the initial parameters~\cite{liu2023analytic}, i.e., when one can safely assume that
\begin{equation}
    \label{eq:linear approx f}
    f_j(\boldsymbol{\theta}_t) = f_j(\boldsymbol{\theta}_0)+ \nabla_\theta f_j(\boldsymbol{\theta}_0)\cdot(\boldsymbol{\theta}_t-\boldsymbol{\theta}_0)\, .
\end{equation}
Here we have introduced the shorthand notation $f_j(\boldsymbol{\theta}_t)\equiv f_j(\mathbf{x}_j,\boldsymbol{\theta}_t)$. Such linearization assumption is valid, for instance, in the so-called \textit{lazy training} regime~\cite{chizat2019lazytraining}. The latter is defined as a condition in which the variational parameters do not change significantly during learning. Notably, this happens in wide (i.e., with many neurons per layer) classical NNs and has been shown to occur in both wide (i.e., with many qubits), shallow ~\cite{Abedi2023quantumlazytraining}, and deep QNNs~\cite{shirai2022quantum, liu2023analytic}. In this work, we will consider the training dynamics to be effectively lazy whenever the relative norm of the parameter updates does not change significantly. 

In general, the QNTK as defined in Eq.~\eqref{eq:QNTK} evolves during training, tracking the current value of the parameters.
However, in the lazy training regime the QNTK becomes \emph{frozen}~\cite{liu2022representation}, i.e., it does not change significantly from its value at initialization throughout the whole training procedure. Under these conditions, one can describe the evolution of the training errors with the following \emph{gradient flow} dynamics
\begin{equation}
\label{eq:error exp decay continuous time}
    \varepsilon_i(t) = \sum_{i'}[(\mathds{1}-\eta K)^t]_{ii'}\varepsilon_{i'}(0) \simeq \sum_{i'}[e^{-\eta Kt}]_{ii'}\varepsilon_{i'}(0) \, ,
\end{equation} 
where the approximation to an exponential decay function is valid for infinitesimal training steps. Combining the definition of  MSE, given in Eq.~\eqref{eq:loss_ntk}, with the time decay of the residual training error, Eq.~\eqref{eq:error exp decay continuous time}, an equation describing the full loss dynamics during training can be formulated as
:
\begin{equation}
    \mathcal{L}(\boldsymbol{\theta})=\frac{1}{2}\sum_{i}\left(\sum_{i'}[e^{-\eta Kt}]_{ii'}\varepsilon_{i'}(0) \right)^2 \, .
\end{equation}
Detailed concentration conditions for the frozen QNTK %
in deep and wide QNNs can be found in Ref.~\cite{liu2023analytic}.

\section{QNTK-based model diagnostics}
\label{sec:diagnostic}

As anticipated in Sec.~\ref{sec:intro}, our goal is to describe how the QNTK {can be leveraged} to gain a-priori insights into the model behavior and its potential performance. To derive our results, we assume that the model operates under lazy training conditions and that the QNKT concentrates around its frozen {average value}. Strictly speaking, this limits the validity of our theoretical analysis to the regime of deep, overparametrized, and wide QNNs as defined in Ref.~\cite{liu2023analytic}. We will later present empirical results which assess the transferability of our results to more realistic and less {constrained} settings. Notably, we find that our theory derived for an idealized setting is in good agreement with the numerical simulations for more realistic models.

Our QTNK-based diagnostic tools rely on the idea that, under the conditions mentioned above, it is possible to approximate the expected model output at an arbitrary training step $t$, for either a \textit{training} or \textit{test} data point $\mathbf{x}_{\gamma}$, using information already available in the training set together with the frozen QNTK matrix evaluated at initialization ($t=0$). In particular, we {derive} the following fundamental result for QNTK diagnostics 
\begin{equation}\label{eq:mean predic in time}
    \mathbb{E}_\theta[f_\gamma( \boldsymbol{\theta}_t)]=\sum_{ii' j} \mathbb{E}_\theta\biggl[\widetilde{K}_{\gamma i}[K^{-1}]_{ii'}\left[\mathds{1}-e^{-\eta K t}\right]_{i' j}\biggr] y_{j}, 
\end{equation}
where all indices $i$, $i'$ and $j$  label \textit{training} data points, $K^{-1}$ is the inverse of the QNTK sub-matrix constructed on the training dataset, and $\widetilde{
K}$ is the natural extension of the QNTK to arbitrary data from $\mathcal{D}$. {The full derivation of the expression in Eq.~\eqref{eq:mean predic in time} is reported in App.~\ref{appendix: integration of param update}}.

From Eq.~\eqref{eq:mean predic in time}, we see that the average model prediction for a generic input $\mathbf{x}_\gamma$ is obtained from a function which only requires information contained in the training dataset $\mathcal{X}$, where the target labels are known, instead of all of $\mathcal{D}$. 
This is done via the linear model mapping
\begin{equation}
    \label{eq:kernel metric}
    \mathcal{M}_{i'}(\mathbf{x}_\gamma)= \sum_{i} \widetilde{K}_{\gamma i}[K^{-1}]_{ii'} \, ,
\end{equation}
which computes the kernel entry $\widetilde{K}_{\gamma i}$ between the new input and all the training samples $\mathbf{x}_{i}$, and then weights the results by the inverse QNTK computed on the training samples alone. %
Note that if $\mathbf{x}_\gamma$ belongs to the training set, then $\mathcal{M}_{i'}(\mathbf{x}_\gamma)=1$ if $\gamma=i'$ and 0 otherwise.

\subsection{Training dynamics}
\label{subsec:training dynamics}

A number of interesting results connecting properties of the QNTK to features of a model's training dynamics can be obtained from Eq.~\eqref{eq:mean predic in time}. To derive them, it is convenient to introduce a slightly different notation. In particular, let us define the vector of mean outputs of the QNN model on the whole training set $\mathcal{X}$ and the corresponding vector of labels as
\begin{equation}
    \label{eq:set formalism}\mathbb{E}_\theta[f(\mathcal{X},\boldsymbol{\theta}_t)]=\begin{bmatrix}
                            \mathbb{E}_\theta[f(\mathbf{x}_0,\boldsymbol{\theta}_t)]\\
                            \vdots\\
                            \mathbb{E}_\theta[f(\mathbf{x}_M,\boldsymbol{\theta}_t)]\\
                            \end{bmatrix}
    , \quad \mathcal{Y} = \begin{bmatrix}
                            y_0\\
                            \vdots\\
                            y_M
                            \end{bmatrix} .
\end{equation}
Here, the $i$-th element of $\mathbb{E}_\theta[f(\mathcal{X},\boldsymbol{\theta}_t)]$ is computed by applying Eq.~\eqref{eq:mean predic in time} to the $i$-th training point:
\begin{equation}
    (\mathbb{E}_\theta[f(\mathcal{X},\boldsymbol{\theta}_t)])_i = \sum_{j} \mathbb{E}_\theta \biggl[\left[\mathds{1}-e^{-\eta K t}\right]_{ij}\biggr] y_{j}\, ,
\end{equation}
where we have used the fact that, by definition, $\widetilde{K}$ and $K^{-1}$ simplify when the former acts on the training set. Notice that the two column vectors introduced above have dimensions $M\times 1$, and therefore belong to the vector space upon which the QNKT acts as a linear operator. Since the latter is symmetric, it can be diagonalized as $K=A^TDA$, where $A$ is an orthogonal matrix and $D=\mathrm{diag}\left(\lambda_0, \ldots \lambda_{M-1}\right)$ contains the eigenvalues $\left\{\lambda_i\right\}_{i\in\left\{0, \ldots, M-1\right\}}$ of $K$. We then have
\begin{equation}
    A(\mathds{1}-\eta K)^tA^T \simeq Ae^{-\eta Kt}A^T=e^{-\eta D t}\simeq (\mathds{1}-\eta D)^t.
\end{equation}
Let us also define the transformed vectors $\widetilde{\mathbb{E}}_\theta[f(\mathcal{X},\boldsymbol{\theta}_t)]=A\mathbb{E}_\theta[f(\mathcal{X},\boldsymbol{\theta}_t)]$ and $\tilde{\mathcal{Y}}=A\mathcal{Y}$ such that
\begin{equation}
    \label{eq:training error eigenvalues}\widetilde{\mathbb{E}}_\theta[f(\mathcal{X},\boldsymbol{\theta}_t)] =  \mathbb{E}_\theta \left[\mathds{1}-e^{-\eta D t}\right]\tilde{\mathcal{Y}} \,,
\end{equation}
with the vector of mean residual training errors 
\begin{equation}
    \widetilde{\mathcal{E}}(t) =\widetilde{\mathbb{E}}_\theta[f(\mathcal{X},\boldsymbol{\theta}_t)] -\tilde{\mathcal{Y}} \simeq -\mathbb{E}_\theta\left[(\mathds{1}-\eta D)^t\right]\tilde{\mathcal{Y}}\,.
\end{equation}
It is now possible to see %
that a necessary condition to ensure that these do not diverge is that
\begin{equation}
  |1-\eta \lambda_{k}|<1 \implies 0<\eta \lambda_{k}<2\,.
\end{equation}
for all eigenvalues $\lambda_k$ of the QNTK~\cite{lewkowycz2020catapult}. Provided that the lazy training/linear approximation regime is valid, the maximum allowed learning rate to guarantee a successful training is therefore 
\begin{equation}
    \label{eq:max_lr_residual}
    \eta_{crit}\approx 2/\lambda_{max},
\end{equation}
where $\lambda_{max}$ is the largest QNTK eigenvalue. Furthermore, once $\eta$ is set to this critical value we get that the slowest component of $\widetilde{\mathcal{E}}(t)$--associated with the smallest eigenvalue $\lambda_{min}$--decays with a characteristic time
\begin{equation}\label{eq:decay rate}
    \tau = (\eta_{crit} \lambda_{min})^{-1}=2\kappa \,,
\end{equation}
where $\kappa=\lambda_{max}/\lambda_{min}$ is the QNTK condition number. This result is closely related to an analogous derivation for classical NNs~\cite{xiao2020disentangling}. It allows us to conclude that $\kappa$--evaluated at initialization--can be used as an informative proxy for the training behavior of a given {QNN} model for a specific dataset.
Specifically, the QNTK condition number can be used to estimate a typical convergence time (controlled by $\lambda_{min}$) when using the largest possible training step (i.e., $2/\lambda_{max}$). 
By taking into account the latter, which is often a crucial hyperparameter in practical scenarios~\cite{moussa2024hyperparameter}, this approach provides a more comprehensive description of the model properties as compared to the sole assessment of the slowest rate of convergence. 
Additionally, $\kappa$ also measures the cost of inverting the QNTK itself, which is a necessary step to calculate the inference map $\mathcal{M}(\mathbf{x})$.

\subsection{Inference}

\begin{figure*}
    \centering
    \includegraphics[trim={.cm .cm .cm .cm},clip,width=1.\textwidth]{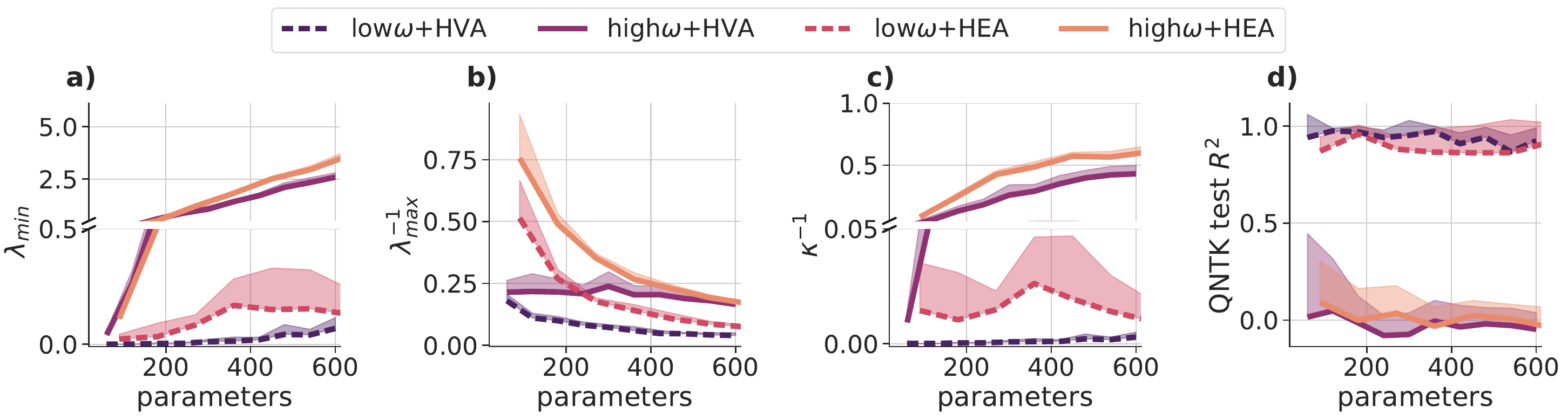}
    \caption{\textbf{QNTK analysis.} This figure summarizes the main features of the QNTK for the TFIM test case as a function of the number of parameters. The smallest \textbf{a)} and largest \textbf{b)} eigenvalues increase ($1/\lambda_{max}$ decreases), as well as the inverse of the condition number $\kappa^{-1}$ (and hence the typical decay rate $\tau^{-1}=2\kappa^{-1}$) \textbf{c)}. Panel \textbf{d)} represent the $R^2$ score of the fit performed by the QNTK over the whole range of $h_X/J_Z$. Lines (solid and dashed) represent the average over 10 different initializations while shaded regions correspond to one standard deviation.}
    \label{fig:TFIM_NTK_summary}
\end{figure*}

Besides assessing how fast a certain QNN model can be trained, {it is typically of interest to have an initial guess on} its potential performance on previously unseen data. In fact, it is well known that a small training error does not directly imply a low \emph{true risk}, which is the value of the loss function evaluated over unseen data $(\mathbf{x}_\gamma, y_\gamma)$ {taken from the $\mathcal{D}$ distribution}. 
For instance, it may happen that the model \emph{overfits}, such that it is particularly good at predicting labels for the training data but does not provide good accuracy for unseen data. This scenario is likely to happen with highly expressive models.
Several strategies have been proposed to mitigate the emergence of overfitting, e.g., by adding suitable inductive biases, constraints or random components~\cite{tian2022regularization,elemstatlear, wan2013dropconnect, srivastava2014dropout, kobayashi2022entangling_drop, scala2023dropout}. However, since most of these techniques need to be applied during the training phase, and often come with additional computational costs, it would be advantageous to determine beforehand whether the model requires them.

Interestingly, also in this context we can leverage QNTK theory to gain a-priori insights. 
Indeed, if we consider the infinite time limit of Eq.~\eqref{eq:mean predic in time}, we find
\begin{equation}
\label{eq:generalization QNTK}
\underset{t\rightarrow\infty}{\mathbb{E}_\theta[f(\mathbf{x}_\gamma,\boldsymbol{\theta}_t)]}=\sum_{i}\mathbb{E}_\theta[\mathcal{M}_{i}(\mathbf{x}_\gamma)]y_{i} \,.
\end{equation}
This suggests that we can use the map $\mathcal{M}_{i}$ from Eq.~\eqref{eq:kernel metric} to predict the model output on unseen test data points.
The calculation is formally exact given the frozen QNTK regime, and--as we will show empirically in the following--it may serve as a good approximation in some practical cases, too. 
Hence, Eq.~\eqref{eq:generalization QNTK} enables us to a-priori estimate test errors and, if necessary, to adapt the model design or change the training scheme, e.g., by applying some regularization method.

It is important to notice that Eq.~\eqref{eq:generalization QNTK} is by construction equivalent to a linear kernel-based model prediction formula~\cite{roberts2022principles}. 
More specifically, it is designed to fit all training data exactly and requires the inversion of the $M\times M$ QNTK matrix.
As a consequence, Eq.~\eqref{eq:generalization QNTK} may fail to capture non-linear contributions with respect to trainable model parameters--similarly to the inability of NTK approximations to describe representation learning capabilities. 
A more general perturbative approach, which goes beyond the frozen QNTK limit, may help to lift some of these theoretical restrictions.

\section{Numerical analysis}
\label{sec:results}

\begin{figure*}
    \centering
    \includegraphics[
    width=1\textwidth]{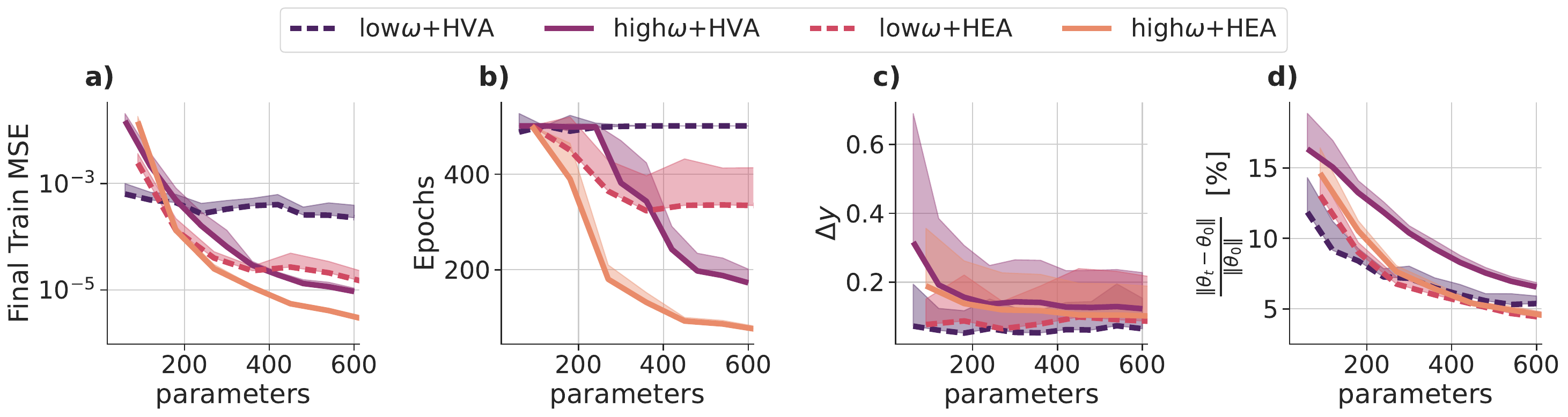}
    \caption{\textbf{Training summary for $\boldsymbol{\eta=\eta_{crit}}$.} Analysis of QNN training on the TFIM dataset. \textbf{a)} Final training MSE. \textbf{b)} Training epochs. \textbf{c)} Average absolute difference between QNN and QNTK-based predictions. \textbf{d)} Relative parameter change during training. Lines (solid and dashed) represent the average over 10 different initializations while shaded regions correspond to one standard deviation.}
    \label{fig:ising_summary_max_lr}
\end{figure*}

In this section, we present numerical experiments performed on concrete choices of QNN models to illustrate the use of the QNTK as a diagnostic tool. 
Descriptions of the employed QNN architectures are provided in App.~\ref{appendix:qnns}.
For the QNN ansatzes, we consider instances of the hardware-efficient ansatz (HEA) and of the Hamiltonian variational ansatz (HVA).
We also vary the data encoding strategy in the QNN models by drawing inspiration from the well-known partial Fourier series description of data re-uploading models~\cite{schuld2021effect}. 
More specifically, we engineer high- and low-frequency data encoding functions, i.e., classes of models with access to Fourier spectra respectively less or more concentrated towards $\omega = 0$. 
We analyze both regression and classification tasks 
on synthetic datasets employing the Mean Squared Error (MSE) loss and vanilla gradient descent (GD) optimization
using the Qiskit~\cite{qiskit2024} statevector simulator. Both the cost function and optimizer choice are consistent with the standard assumptions of QNTK theory, see Sec.~\ref{sec:background}. Unless otherwise stated, all trainings are carried out with the critical learning rate $\eta_{crit}$.
Additional results on different are presented in App.~\ref{appendix: other results}.

As a representative example, we investigate the prediction of the ground state transverse magnetization $\langle\sigma_X\rangle$ for a 1D transverse field Ising model (TFIM) with periodic boundary conditions as a function of the external field. 
The 1D TFIM describes a spin lattice 
characterized by nearest-neighbour interactions along the $Z$ axis and by an external magnetic field $h_X$ along the $X$ axis. 
This system is described by the following Hamiltonian:
\begin{equation}
    H_k=-J_Z\sum_{i=1}^N \sigma_Z^{i} \sigma_Z^{i+1} - h^k_X \sum_{i=1}^N \sigma_X^{i} \,.
\end{equation}
To construct our dataset, we choose $20$ different values of the external magnetic field such that $-5 \leq h^k_X /J_Z \leq4.5$.
The dataset consists of approximate ground state transverse magnetization $\langle\sigma^k_X\rangle$ values as a function of the external field, for a system of $N=6$ spins. 
To compute them, we first find
an approximate ground state $\ket{\psi(\vartheta_k^*)}$ for each of the $20$ different values of $h^k_X$ using the Variational Quantum Eigensolver (VQE)~\cite{Peruzzo2014VQE}algorithm. We employ a 
Hamiltonian Variational Ansatz (HVA), which is known to be well-suited for this kind of application, built with layers of parametrized single qubit rotations 
around the $X$ axis on each qubit and $R_{ZZ}$ parametrized 
entangling rotations connecting neighbouring qubits (with periodic boundary conditions).
Using COBYLA~\cite{powell1998cobyla}, we optimize a layout with 25 HVA layers to obtain 
the the optimal parameters $\vartheta_k^*$ to approximate the exact ground states.
We then compute the transverse magnetization as $\langle\sigma^k_X\rangle=\bra{\psi(\vartheta_k^*)}\sigma^k_X\ket{\psi(\vartheta_k^*)}$. 
In this way, our TFIM dataset comprises 20 points with $h^k_X$ corresponding to the input data and $\langle\sigma^k_X\rangle$ to the respective labels, $\left\{{h^k_X}, \langle\sigma^k_X\rangle_i\right\}_{k=1}^{20}$. We apply a random split in training and test set with a $50\%$ ratio, as shown in Fig.~\ref{fig:datasets} of Appendix~\ref{appendix: other results}. All the target expectation values lie in $[-0.8,0.8]\subset [-1,1]$, so we do not apply any rescaling. %
Instead, we pre-process the input variables $h^k_X/J_Z$ with a min-max scaler from  the \texttt{sklearn} Python library~\cite{sklearn}, mapping the features in the interval $[-0.95, 0.95]$.

Inference capabilities are assessed on the test data by computing the coefficient of determination~\cite{r2score}, defined as
\begin{equation}
\label{eq:R2}
    R^2=1-\frac{\sum_i \varepsilon_i^2}{\sum_i (y_i-\Bar{y})^2}\, ,
\end{equation}
where $\Bar{y}_i=\sum_{j=1}^M y_j/M$ is the mean value of the data labels. 
This is an established metric for regression tasks. Optimal fitting 
leads to $R^2=1$ and $R^2<1$ indicates worse performance (negative values are also allowed). \\
In Fig.~\ref{fig:TFIM_NTK_summary}, we show how the average QNTK spectral properties evolve for progressively deeper circuits. Specifically, Fig.~\ref{fig:TFIM_NTK_summary}a displays the growth of the smallest eigenvalue $\lambda_{min}$ as more trainable layers, and therefore more trainable parameters, are added to the QNN ansatz. As discussed in Sec.~\ref{subsec:training dynamics}, $\lambda_{min}$ is related to the speed of convergence during training, with larger minimum eigenvalues implying faster training at constant learning rates. It is worth mentioning that such an increase of $\lambda_{min}$ for deeper QNNs is in good agreement with known theoretical results on training in overparametrized models~\cite{kiani2020learning,larocca2023theory}. In practice, our results suggest that QNNs with access to higher encoding frequencies will show faster training convergence compared to low-frequency ones. Furthermore, we expect that HEA-based models will outperform HVA-based ones in terms of training speed.
Fig.~\ref{fig:TFIM_NTK_summary}b indicates that $\lambda_{max}$ increases with the number of parameters, such that the critical learning rate $\eta_{crit}\propto \lambda_{max}^{-1}$ must progressively be reduced. It should further be noted that the plot shows that $\eta_{crit}$ remains larger for high-frequency and HEA-based models compared to low-frequency and HVA-based ones.

As shown in Fig.~\ref{fig:TFIM_NTK_summary}c, the overall tendency of the QNTK condition number $\kappa$ is to predominantly decay for deeper models, as $\lambda_{min}$ rises faster than $\lambda_{max}$. This behavior suggests a generic speed up in the training process. 
Interestingly, the decrease in the QNTK condition number is much stronger for QNNs with high-frequency encodings than for low-frequency ones. 
Small condition numbers imply fast convergence in the training error. Hence, the respective systems can be prone to overfitting.
This is essentially confirmed by Fig.~\ref{fig:TFIM_NTK_summary}d, which illustrates an assessment of the inference capabilities obtained by measuring the $R^2$ score on the predictions from Eq.~\eqref{eq:generalization QNTK}. 
Based on our QNTK diagnostics, we therefore expect that the high-frequency QNNs are prone to overfitting the training data. Some form of regularization such as an early stopping approach, could then be beneficial when dealing with this class of models. Notice that our conclusions are in line with insights presented in the pioneering work of Jacot et al.~\cite{jacot2018ntk}. \\

\begin{figure*}
    \centering
    \includegraphics[width=1.\linewidth]{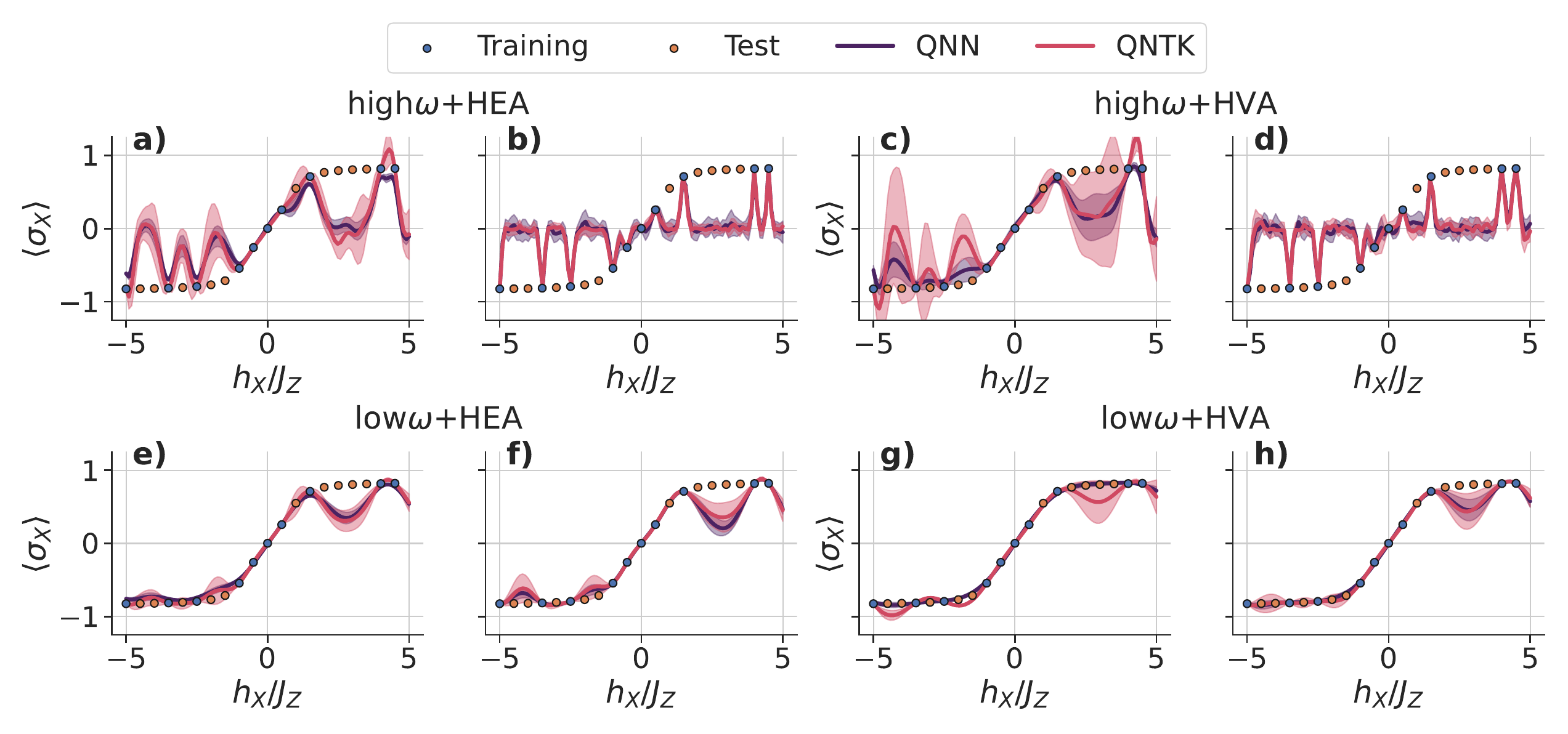}
    \caption{\textbf{Inference summary for $\boldsymbol{\eta=\eta_{crit}}$.} For each QNN model design, two panels compare the QNTK inference approximation (red line)  with its trained QNN counterpart (purple line) over the 1D TFIM dataset at 5 (left sub-plot) and 50 (right sub-plot) layers. Specifically, we have: high-$\omega$ with HEA using \textbf{a)} 5 and \textbf{b)} 50 layers, high-$\omega$ with HVA using \textbf{c)} 5 and \textbf{d)} 50 layers, low-$\omega$ with HEA using \textbf{e)} 5 and \textbf{f)} 50 layers, low-$\omega$ with HVA using \textbf{g)} 5 and \textbf{h)} 50 layers. Lines represent the average over 10 different initializations while shaded regions correspond to one standard deviation. Blue points denote training data, and orange ones are test samples.}
    \label{fig:5_vs_50}
\end{figure*}

It is now interesting to compare our diagnostic predictions based on the QNTK analysis with the actual behavior of the QNN models once trained. To carry out the learning procedure, we used a learning rate $\eta\leq\eta_{crit}$ and a maximal number of iterations equal to 500. In some instances, we also tested the models for instabilities using $\eta>\eta_{crit}$ with a maximum of 250 iterations.  Unless specified otherwise, the convergence criterion on the trainable parameters was chosen as
\begin{equation}
\label{eq:convergence crit}
    \|\boldsymbol{\theta}(t+1)-\boldsymbol{\theta}(t)\|<10^{-3}\,.
\end{equation}

Let us first discuss the results for the actual QNN models with up to 50 layers concerning the training dynamics at $\eta=\eta_{crit}$.  In Fig.~\ref{fig:ising_summary_max_lr}a-b we report the final training MSE and the number of epochs averaged over 10 different random initializations. In agreement with our QNTK-based study, both metrics consistently decrease, i.e., the training becomes shorter, as the models grow deeper--except for the low-$\omega$/HVA case, which always reaches the maximum allowed number of iterations. 
Notably, high-frequency models are the fastest to reach convergence and even lead to the lowest errors. Furthermore, in Fig.~\ref{fig:ising_summary_max_lr}c, we describe how well the QNTK-based inference predictions from Eq.~\eqref{eq:generalization QNTK} approximate the actual QNN ones. 
This is measured via the Average Absolute Difference (AAD), 
\begin{equation}
\label{eq:aad}
    \Delta y = \frac{1}{D}\sum_i^D\left|y_i^{QNN}-y_i^{QNTK}\right| \, ,
\end{equation}
where $y_i^{QNN}$ and $y_i^{QNTK}$ are the vectors of predictions produced by the trained QNN model and the QNTK map. 
In order to investigate in depth the quality of the approximation, we evaluate the AAD on an extended test set composed of $D=100$ uniformly distributed input samples over the whole range of $h_X/J_Z$. 
The results indicate that, as the number of parameters increases, the AAD diminishes to approximately 0, thus implying that--consistently with the general theory--the QNTK-based predictions become more and more accurate as we enter the deep overparametrized regime. 
As a consequence, we expect that the overfitting behavior of high-frequency models foreseen in Fig.~\ref{fig:TFIM_NTK_summary} will indeed take place in this regime. 
Lastly, in Fig.~\ref{fig:ising_summary_max_lr}d we check the validity of the lazy training assumption by measuring the average relative change in the parameters from initialization to the last epoch. 
By comparing Figs.~\ref{fig:ising_summary_max_lr}c and~\ref{fig:ising_summary_max_lr}d, we then notice that when the parameters change more than $10\%$ the QNTK-based approximations tend to be less accurate, as the lazy training assumption is effectively violated.

To offer a concrete visualization of the results, we plot the actual QNN (purple line) and the QNTK-based (red line) predictions for 5- and 50-layer models in Fig.~\ref{fig:5_vs_50}. By comparing the different panels, one can immediately appreciate the distinction between the high- and low-frequency cases, with the former being able to express very fast oscillations and--as a consequence--being very prone to overfitting the training data (blue points) while poorly predicting most of the test ones (orange points). This behavior becomes even more evident as the number of layers grow, since the models get access to a larger spectrum of frequencies (see App.~\ref{appendix:qnns}). 

In all cases, the QNTK-based predictions are in excellent agreement with the QNN results. Interestingly, one can clearly observe that, as the number of layers increase, the QNTK concentrates more and more around its mean initialization value~\cite{liu2023analytic}, and the standard deviation of QNTK predictions becomes smaller. At the same time, the QNTK-based model offers a good approximation already for 5-layer models: this may suggest that, at least under certain conditions, the regime of validity for QNTK diagnostics might be extended, perhaps in a perturbative way, to relatively shallow QNNs. %

\begin{figure*}
    \centering
    \includegraphics[
    width=1\textwidth]{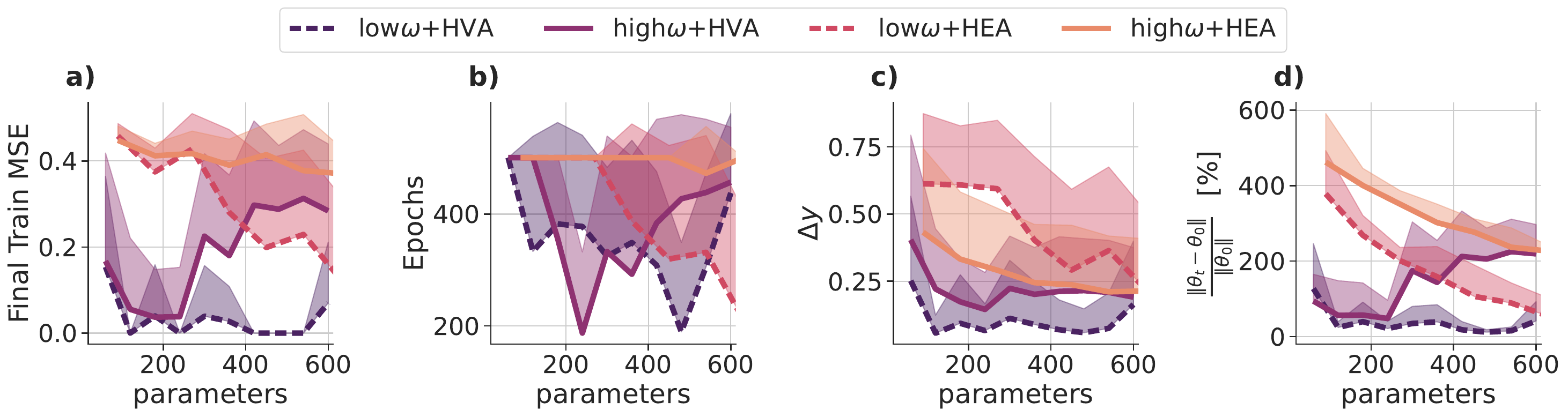}
    \caption{\textbf{Training summary for $\boldsymbol{\eta=10\eta_{crit}}$.} Analysis of QNN training on the TFIM dataset above the critical learning rate. \textbf{a)} Final training MSE. \textbf{b)} Training epochs. \textbf{c)} Average absolute difference between QNN and QNTK-based predictions. \textbf{d)} Relative parameter change during training. Lines (solid and dashed) represent the average over 10 different initializations while shaded regions correspond to one standard deviation.}
    \label{fig:ising_summary_10_max_lr}
\end{figure*}

Lastly, in Fig.~\ref{fig:ising_summary_10_max_lr} we test the possible disruptive effects of setting the learning rate beyond the critical value determined from the QNTK eigenspectrum, see Eq.~\eqref{eq:max_lr_residual}. 
More specifically, we perform a training experiment with $\eta=10\eta_{crit}$. In such extreme conditions, severe oscillations around the loss landscape can be expected and are indeed confirmed, as shown in  Fig.~\ref{fig:ising_summary_10_max_lr}a and~\ref{fig:ising_summary_10_max_lr}b. 
Additionally, the relative parameter change across the whole training procedure is always above $10\%$ and can reach values of the order of hundreds, see Fig.~\ref{fig:ising_summary_10_max_lr}d. This means that all models operate well beyond the lazy training regime, with the AAD in Fig.~\ref{fig:ising_summary_10_max_lr}c achieving non-negligible values. 
Interestingly, the low-frequency and HVA-based QNNs appear to be slightly more resilient compared to other choices of model architecture.

\section{Conclusions and outlook}
\label{sec:conclusions}

Inspired by the use of NTKs in classical ML, this work explored the use of the QNKT as a diagnostic tool for quantum learning models based on parametrized quantum circuits. 
In particular, we derived analytical results connecting spectral features of the QNKT, such as the maximum eigenvalue and the condition number computed at model initialization, with important qualitative features of the training dynamics, identifying a critical learning rate and a characteristic decay time for average residual errors.
Furthermore, we obtained a QNTK-based linear model approximation that allows us to predict the expected inference output of a quantum neural network. 
Crucially, our protocol is resource efficient as it only requires one gradient evaluation per training or test point and at most the inversion of a kernel matrix with dimensions running over the size of the dataset. 
We validated our predictions on a variety of model classes, showcasing how our proposed techniques can help assessing the dataset-dependent performances of a given QNN, and to identify potential pitfalls in model designs. 

Strictly speaking, our formal analysis only holds for averaged training errors and model predictions, and in a rather idealised setting, i.e., a lazy training regime with frozen QNTK. In other words, all of our conclusions derive from a linear approximation of the average QNN dynamics. As we already mentioned in the main text, this results in a few potential limitations. Specifically, 
our QNTK-based models exactly fit, by definition, all training points, while this is not necessarily true for the corresponding QNN after training. As a consequence, the QNTK might in practice only {be useful} as a rough estimate of the generalization properties of the model under study.
For instance, in relatively shallow models we observed the tendency of the QNTK matrix to become singular. 
Nevertheless, whenever the QNTK remains well-conditioned and the QNN itself is expressive enough to converge to a global minimum of the training loss--conditions that are easily satisfied when models operate not too far away from a lazy training regime--we empirically find good diagnostic predictions even for shallow trainable ansatzes. A deeper understanding of this behavior would require a full perturbative analysis, including higher order approximations starting from a running QNTK with a frozen second-order meta-kernel~\cite{roberts2022principles,liu2022representation}, up to a more detailed investigation of the full single-trajectory dynamics~\cite{you2023analyzing}.
In this context, it would also be interesting to investigate in more detail the relationship between the typical signatures of overparametrization and the emergence of barren plateaus, as characterized by different mathematical tools such as, e.g., the QNTK, dynamical Lie algebras~\cite{larocca2023theory}, or the quantum Fisher information matrix~\cite{abbas2021power}. 

\section*{Acknowledgments}
The authors warmly acknowledge Antonio Mezzacapo and Kunal Sharma for fruitful discussions and constructive feedback.
{This work was partly supported by the Italian Ministry of University and Research (MUR) after 
funding from the European Union – NextGenerationEU through the National Recovery and Resilience Plan (NRRP) with the “National Quantum Science and Technology Institute” (NQSTI, Grant No. PE0000023). F.S. acknowledges the Erasmus Traineeship program of the University of Pavia for partial support during his research visit at IBM Research Zurich.} IBM, the IBM logo, and ibm.com are trademarks of International Business Machines Corp., registered in many jurisdictions worldwide. Other product and service names might be trademarks of IBM or other companies. The current list of IBM trademarks is available at \url{https://www.ibm.com/legal/copytrade.}

\bibliography{bibliography}%

\newpage
\appendix
\onecolumngrid

\section{The Neural Tangent Kernel}%
\label{appendix:NTK}
In this section, %
we sketch for completeness a derivation of the classical NTK for arbitrary input and output dimensions. %
In a supervised ML framework, one typically employs a parametrized model ${f}(\mathbf{x}, \boldsymbol{\theta})$ to extract and refine information from a given dataset $\mathcal{X}\subset\mathbb{R}^N$ so that, after training, it is possible to predict $C$-dimensional properties (labels) for some previously unseen data
\begin{align}
    f:& \mathcal{X}\times \mathbb{R}^P\rightarrow \mathbb{R}^C \\
    & (\mathbf{x}_i, \boldsymbol{\theta})\mapsto f (\mathbf{x}_i, \boldsymbol{\theta})=\mathbf{y}_i \quad .
\end{align}
To achieve this goal one usually defines a loss function to be minimized by changing the model parameters. In our case, we will employ the Mean Squared Error (MSE)
\begin{align}
\label{eq:loss_ntk_appendix}
    \mathcal{L(\boldsymbol{\theta})}&= \frac{1}{2}\sum _{i=1}^M\sum_{j=1}^C(f_{j}(\mathbf{x}_i,\boldsymbol{\theta})-y_{ij})^2\\
    &=\frac{1}{2}\sum _{i,j}^{M,C}(f_{ij}(\boldsymbol{\theta})-y_{ij})^2=\\
    &=\frac{1}{2}\sum _{\alpha}^{MC}(f_{\alpha}(\boldsymbol{\theta})-y_{\alpha})^2=\frac{1}{2}\sum _{\alpha}^{MC}\varepsilon_{\alpha}^2
\end{align}
where $\mathbf{x}_i$ is an input among the $M$ possible in the training dataset $\mathcal{X}$, $j$ represents a specific output component and $\varepsilon$ denotes the residual training error. For ease of notation, we merged the input and output indexes into a single one, $(i,j)\rightarrow \alpha$. Notice that this notation is employed only in this section to present the most general framework, whereas, in the rest of this work, we will treat the simpler case of $C=1$.

Classical artificial NNs (ANNs) are a specific class of parametrized models, composed of simple linear units named neurons, which are nonlinearly combined in order to approximate larger, parametrized nonlinear functions. The microscopic behavior of the single units is well understood, but as we put together lots of these elements the dynamics of ANNs may drastically change. For this reason, following an approach inspired by statistical mechanics, ANNs have been studied in various limits to capture their average behaviour during training~\cite{decelle2022introduction}. Of particular interest is the infinite-width limit, in which an ANN has infinitely many neurons in each layer. In this setting, things greatly simplify, as all the derivatives (with respect to the parameters) of order higher than 2 vanish and the training dynamics becomes \emph{linear}~\cite{roberts2022principles}. The ANN then enters the so-called \emph{lazy training} regime~\cite{chizat2019lazytraining}, a condition in which the parameters do not change significantly during training and remain close to their initial values. 
This motivates the study of the linear approximation (in the parameters) of the model function ${f}(\mathbf{x}, \boldsymbol{\theta})$ around the initial parameters through a Taylor series expansion. \\

As anticipated, the goal  of training is to find the minimum of a loss function, such as the MSE. To achieve this, a very common strategy in ML is to perform the update of the parameters in the opposite direction of the gradient of the loss function. This method takes the name of Gradient Descent (GD) and. Applying GD to the MSE in Eq.~\eqref{eq:loss_ntk_appendix} yields the following updated rule for the variational parameters:
\begin{align}
    \delta\theta_l  &= \theta_l(t+1)-\theta_l(t)=\\ 
    \label{eq:param_update_appendix}
    &=-\eta \frac{\partial \mathcal{L}}{\partial\theta_l}= -\eta \sum _{\alpha'}\varepsilon_{\alpha'}\frac{\partial \varepsilon_{\alpha'}}{\partial\theta_l}\, .
\end{align}
Here, the constant $\eta$ is the so-called learning rate and $t$ refers to the iterative step of the gradient descent dynamics. Similarly, we define the change in the residual training error as
\begin{align}
\label{eq:training error_appendix}
    \delta \varepsilon_{\alpha} &= \varepsilon_{\alpha}(t+1)-\varepsilon_{\alpha}(t)= \\
    &=\sum_l \frac{\partial \varepsilon_{\alpha}}{\partial\theta_l} \delta \theta_l \, .
\end{align}
Inserting $\delta\theta_l$ from Eq.~\eqref{eq:param_update_appendix} and re-elaborating a bit, we get
\begin{align}
\delta \varepsilon_{\alpha} &=-\eta \sum _{\alpha'}\varepsilon_{\alpha'}\sum_l \frac{\partial \varepsilon_{\alpha}}{\partial\theta_l} \frac{\partial \varepsilon_{\alpha'}}{\partial\theta_l} = \nonumber\\
    &=-\eta \sum _{\alpha'} K_{\alpha\alpha'}\varepsilon_{\alpha'}
\end{align}
where
\begin{equation}
    \label{eq:NTK}
    K_{\alpha\alpha'}=\sum_l \frac{\partial \varepsilon_{\alpha}}{\partial\theta_l} \frac{\partial \varepsilon_{\alpha'}}{\partial\theta_l}
\end{equation}
is called the Neural Tangent Kernel (NTK)~\cite{jacot2018ntk, liu2022representation, liu2023analytic, shirai2022quantum, roberts2022principles}, which is a symmetric positive semidefinite matrix of dimension $ MC\times MC$. The dimensions of rows and columns are given by the number of inputs and outputs of the parametrized model. One can also go beyond the frozen dynamics approximation by employing the so-called \emph{differential} NTK, which introduces some degree of nonlinearity and allows the NTK to change over time~\cite{roberts2022principles,liu2022representation}. %

\section{Average network prediction formula}
\label{appendix: integration of param update}
In this section, we present the full derivation of Eq.~\eqref{eq:mean predic in time}, which describes the average network prediction during training under a gradient flow dynamics
\begin{equation}
    \mathbb{E}_\theta[f_\gamma( \boldsymbol{\theta}_t)]=\sum_{ii' j} \mathbb{E}_\theta\biggl[\widetilde{K}_{\gamma i}[K^{-1}]_{ii'}\left[\mathds{1}-e^{-\eta K t}\right]_{i' j}\biggr] y_{j} \,.
\end{equation}
Crucially, this result underpins the two main equations used in the main text, i.e., Eq.~\eqref{eq:training error eigenvalues} and Eq.~\eqref{eq:generalization QNTK}.

As the number of epochs in a training process is usually high, we can think of every step as infinitesimally small and switch to the gradient flow picture, i.e., the continuous-time limit, such that we can proceed analogously to the derivation of Eq.~\eqref{eq:error exp decay continuous time}. We can then write, in the linear regime, the following differential equation %
\begin{align}
    \frac{d \theta_l(t)}{d t}=-\eta  \frac{\partial\mathcal{L}(\boldsymbol{\theta}(t))}{\partial \theta_l}=-\eta \sum_{i}  \frac{\partial\varepsilon_{i}(t)}{\partial \theta_l}\varepsilon_{i}(t)=-\eta \sum_{i}  \frac{\partial\varepsilon_{i}(0)}{\partial \theta_l}\varepsilon_{i}(t)\, ,
\end{align}
where we made the $\varepsilon_{i}$ dependence of $ \partial\mathcal{L}/\partial\theta$ explicit. Here the lazy training regime assumption implies linear evolution of the model with respect to the variational parameters, i.e. $ \partial f/\partial\theta$ can be considered constant at all times $t$. As $\varepsilon_i= f(x_i, \boldsymbol{\theta})-y_i$, it follows that also $ \partial \varepsilon/\partial\theta$ can be considered constant and consequently we can evaluate it a $t=0$.

After training for a time $t$, the total change in the parameters is therefore
\begin{align}
\theta_l(t)-\theta_l(0)&= \int_0^{t} d \theta_l(t)=\\
&= -\eta \int_0^{t} \sum_{i} \frac{\partial\varepsilon_{i}(0)}{\partial \theta_l}\varepsilon_{i}(t')dt'=
\\
&=-\eta \int_0^{t} \sum_{i} \frac{\partial\varepsilon_{i}(0)}{\partial \theta_l} \sum_{j}[e^{-\eta Kt'}]_{ij}\varepsilon_{j}(0)dt'=\\
&=-\eta \sum_{i} \frac{\partial\varepsilon_{i}(0)}{\partial \theta_l}\sum_{j}\biggl(\int_0^{t}[e^{-\eta Kt'}]_{ij}dt'\biggr)\varepsilon_{j}(0)=\\
&=-\eta \sum_{i} \frac{\partial\varepsilon_{i}(0)}{\partial \theta_l}\sum_{j}\biggl([K^{-1}]_{ii'}[\mathds{1}-e^{-\eta Kt'}]_{i'j}dt'\biggr)\varepsilon_{j}(0)=\\
\label{eq:delta_theta_appendix}
&=- \sum_{ii'j}  \frac{\partial\varepsilon_{i}(0)}{\partial \theta_l} [K^{-1}]_{ii'}\left[\mathds{1}-e^{-\eta K t}\right]_{i'j}\varepsilon_{j}(0) \, ,
\end{align}
where the indices $i,i',j$ denote data points from the \textit{training} dataset. Notice that we exploited Eq.~\eqref{eq:error exp decay continuous time} in the second line. In vectorial notation, this reads
\begin{align}
    \boldsymbol{\theta}_t-\boldsymbol{\theta}_0&=- \sum_{ii'j}  \nabla_{\theta} \varepsilon_{i} [K^{-1}]_{ii'}\left[\mathds{1}-e^{-\eta K t}\right]_{i'j}\varepsilon_{j}(0)=\\
    \label{eq:delta_theta}
    &=- \sum_{ii'j}  \nabla_{\theta} \varepsilon_{i} [K^{-1}]_{ii'}\left[\mathds{1}-e^{-\eta K t}\right]_{i'j}\left(f_{j}(\boldsymbol{\theta}_0)-y_{j}\right)\, .
\end{align}
By substituting this parameter difference ($\boldsymbol{\theta}_t-\boldsymbol{\theta}_0$) into Eq.~\eqref{eq:linear approx f} and averaging over the distribution of initial parameters, one can approximate the expected model output at training step $t$ for a \textit{training} or \textit{test} data point $\mathbf{x}_{\gamma}$ via the frozen QNTK at initialization ($t=0$). 
In particular, we have
\begin{align}
    \mathbb{E}_\theta[f_\gamma( \boldsymbol{\theta}_t)]&=\mathbb{E}_\theta[f_\gamma( \boldsymbol{\theta}_0)]+ \mathbb{E}_\theta[\nabla_\theta f_\gamma(\boldsymbol{\theta}_0)\cdot\boldsymbol{\theta}_t-\boldsymbol{\theta}_0]=
    \\
    &= - \sum_{ii' j} \mathbb{E}_\theta\biggl[ \nabla_\theta f_{\gamma}   \nabla_{\theta} \varepsilon_{i} [K^{-1}]_{ii'}\left[\mathds{1}-e^{-\eta K t}\right]_{i'j}\biggr]\mathbb{E}_\theta[\varepsilon_{j}(0)]=
    \\
    &= - \sum_{ii' j}  \mathbb{E}_\theta\biggl[\widetilde{K}_{\gamma i}[K^{-1}]_{ii'}\left[\mathds{1}-e^{-\eta K t}\right]_{i'j}\biggr]\left[\mathbb{E}_\theta[f_{j}( \boldsymbol{\theta}_0)]-y_{j}\right]=
    \\
    \label{eq:mean predic in time appendix}
    &=\sum_{ii' j} \mathbb{E}_\theta\biggl[\widetilde{K}_{\gamma i}[K^{-1}]_{ii'}\left[\mathds{1}-e^{-\eta K t}\right]_{i' j}\biggr] y_{j}\,.
\end{align}
Here, we assumed that the expected output of the model over the initial parameter distribution is zero (i.e., $\mathbb{E}_\theta[f(\mathbf{x},\boldsymbol{\theta}_0)]=0$) for any $\mathbf{x}\in \mathcal{D}$. In fact, it was shown in Refs.~\cite{garcíamartín2023deep,girardi2024trained} that the outputs of deep and wide QNNs form Gaussian processes with zero mean with respect to the parameters, and that, asymptotically, the resulting distribution has moments matching Gaussian ones. We then substituted Eq.~\eqref{eq:delta_theta}, recalled that $\nabla_{\boldsymbol{\theta}} f _{\gamma}=\nabla_{\boldsymbol{\theta}} \varepsilon_{\gamma}$ and defined the kernel element $\widetilde{K}_{\gamma i}=\nabla_\theta \varepsilon_{\gamma} \nabla_\theta \varepsilon_{i}$. As a last step, we applied again $\mathbb{E}_\theta[f(\mathbf{x},\boldsymbol{\theta}_0)]=0$. %

\section{Quantum Neural Network Architectures}
\label{appendix:qnns}

\begin{figure}[ht!]
    \centering
    \includegraphics[trim={8.5cm 3.5cm 8.5cm 3.5cm},clip,width=.75\linewidth]{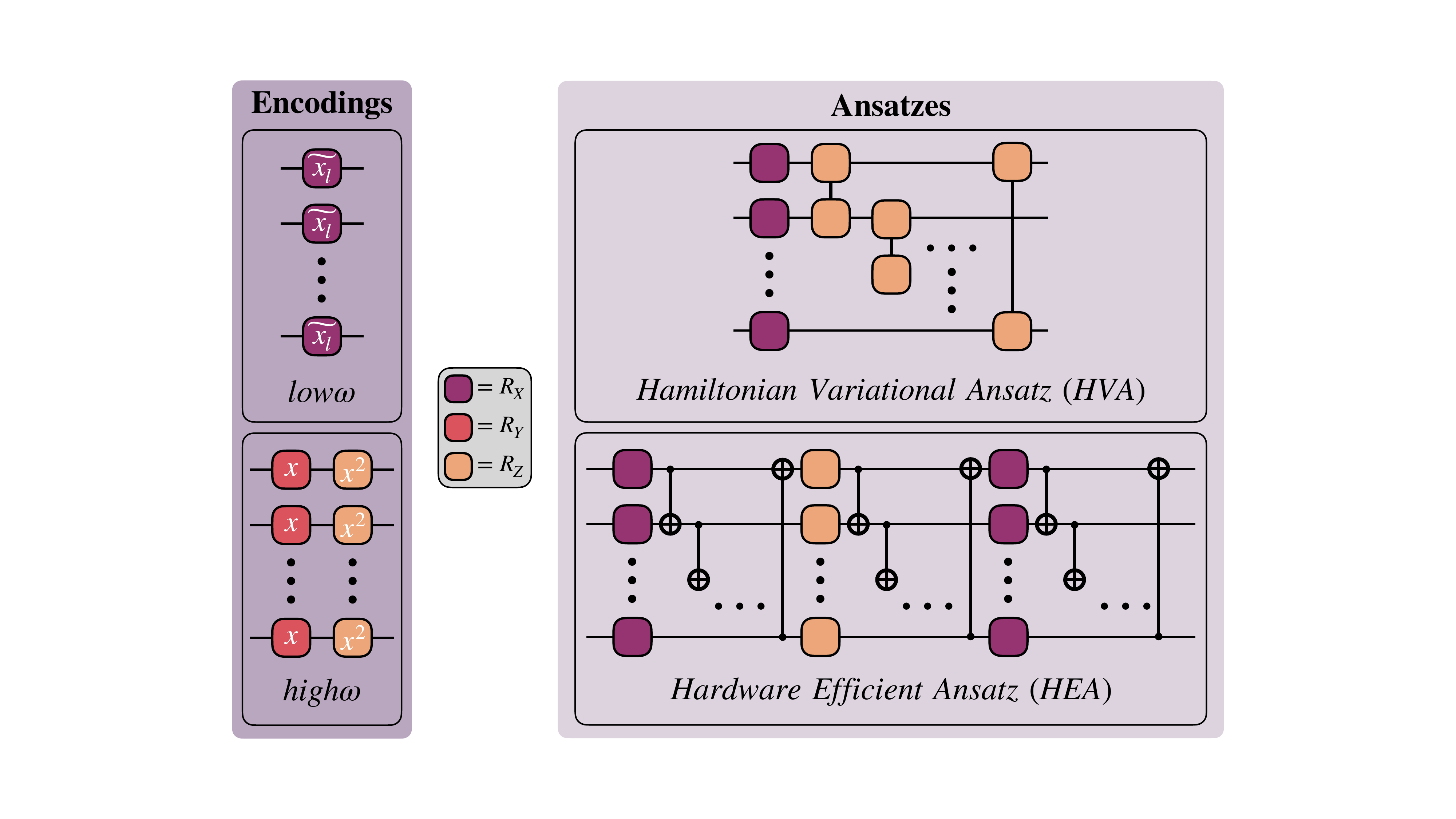}
    \caption{Sketch of the encoding strategies (left) and parametrized ansates (right) employed in this work. In the parametrized ansates we implicitly assume the parameter of a rotational gate to be $\theta$.}
    \label{fig:qnn_architectures}
\end{figure}

In this work we employ four different kinds of QNNs by combining two encoding strategies $S_l(\mathbf{x})$, dubbed low- (low$\omega$) and high-frequency (high$\omega$) encoding, and two variational ansatzes $U_l(\boldsymbol{\theta})$, namely the Hardware Efficient Ansatz (HEA)~\cite{leone2024HEA} and the Hamiltonian Variational Ansatz (HVA). A schematic representation of these circuit elements can be found in Fig.~\ref{fig:qnn_architectures}.

Both encoding strategies are Pauli embeddings with data re-uploading (on parallel qubits)~\cite{schuld2021effect} and classical pre-processing. 
They differ in the type and number of encoding gates and also in the classical pre-processing.
Standard angle embedding uses each component of the data as an angle for parametrized Pauli rotations
\begin{equation}
\label{eq:angle_embedding}
    S_j([\mathbf{x}]_j)=e^{-iP_j[\mathbf{x}]_j}
\end{equation}
where $P_j\in\{\sigma_X,\sigma_Y,\sigma_Z\}$. 
When the dimension of the input is bigger then one, we encode each component $[\mathbf{x}]_j$ one at a time on different qubits. If the number of qubits exceeds the number of components then we repeat this scheme until each qubit has an encoded component.  In particular, in this work we designed two different angle encoding strategies that differ for the chosen encoding gates and classical pre-processing function $g(x)$ providing extra nonlinearity. The first one, referred as the \emph{low-frequency encoding}, employs a single-layer of $R_X$ rotations and pre-processing $g^{(l)}_1(x)$ of the input depending on the layer, where the feature is encoded as:
\begin{equation}
\label{eq:widetildex_l}
 \widetilde{x}_l=g^{(l)}_1(x)=\frac{x}{L-l}   
\end{equation}
where $L$ is the total number of layers and $l$ is the layer index. This strategy aims at lowering the Fourier frequency of a factor $L-l$, giving rise to many different frequencies all concentrated around zero.
On the other hand, we define the \emph{high-frequency embedding} employing a single layer of $R_Y$ rotations followed by another layer of $R_Z$ rotations. The two layers of rotations differ in the pre-processing: $R_Y$ trivially encode the input itself $g_2(x)=x$, while $R_Z$ encodes $g_3(x)=x^2$. We stress that $g_2$ and $g_3$ are layer independent. One can better understand the role of such pre-processing functions by looking at Eq.~\eqref{eq:fourier}. \\

The embedding is followed by the variational ansatz $U_l(\boldsymbol{\theta})$.  
The HEA is composed of three sub-layers with single qubit rotations, followed by CNOT gates connected in a ring shape which is equivalent to linear nearest-neighbour connectivity with periodical boundary conditions. The first and third sub-layers use $R_X$ rotations whereas the second applies $R_Z$ gates. The HVA is modeled w.r.t.~a 1-dimensional Ising model with a transverse field on the $X$ axis. 
Hence, we have single qubit $R_X$ rotations followed by $R_{ZZ}$ entangling gates that connect qubits with nearest-neighbour connectivity and periodical boundary conditions.

\begin{figure*}
    \centering
    \includegraphics[width=\linewidth]{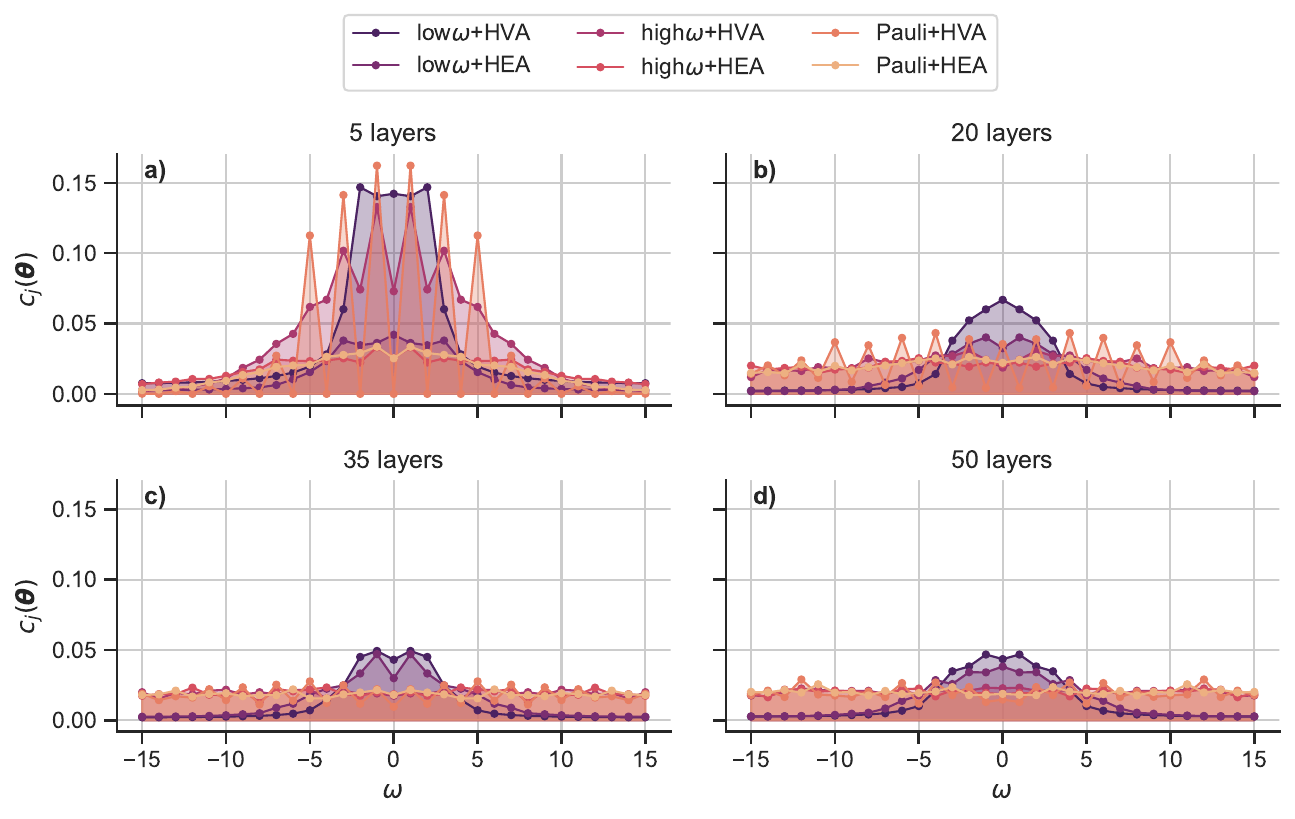}
    \caption{Instensity of Fourier coefficients in for different QNN architectures and different number of layers: \textbf{a)} 5, \textbf{b)} 20, \textbf{c)} 35 and \textbf{d)} 50. The spectrum is truncated.}
    \label{fig:fourier}
\end{figure*}

Data encoding layers and variational ones can be repeated multiple times, following what is called a \emph{data re-uploading} strategy~\cite{perez2020data, gil2020input, schuld2021fourier,nemkov2023fourier}. 
This not only enhances the expressiveness of the model but also opens up the possibility of writing the model in terms of a truncated Fourier series. More specifically, for a given classical input $\mathbf{x}\in\mathcal{X}$
\begin{equation}
\label{eq:fourier}
    f(\mathbf{x}_i, \boldsymbol{\theta})=\langle\psi(\mathbf{x}_i, \boldsymbol{\theta})| O \ket{\psi(\mathbf{x}_i, \boldsymbol{\theta})}= \sum_{j\in \Lambda}c_j(\boldsymbol{\theta})e^{i\boldsymbol\omega_j \cdot g(\mathbf{x}_i)},
\end{equation}
where $\Lambda=\{j\in \mathbb{Z}|\omega_j \in \Omega\}$ with $\Omega$ being the Fourier spectrum and $g$ is a pre-processing function. 
Note that only the coefficients $c_j(\boldsymbol{\theta})$ depend on the parametrized operations (as well as on the constant gates in the training layers), such that the training process cannot affect the available frequencies but may change their contribution. 
Depending on the intensity of the coefficients, the available frequencies in the spectrum $\Omega$ can then contribute differently to the computation. We define low/high-frequency models precisely as the ones favoring the relevant parts of the Fourier spectrum.

We verify the effect of our encoding strategies numerically. The respective results are shown in Fig.~\ref{fig:fourier}.
The plot illustrates the average value of the Fourier coefficients associated to each frequency of the (truncated) spectrum over 20 different random weight vectors, measuring the Pauli $Z$ operator on the first qubit. 
The coefficients are computed with the fast Fourier transform given by Numpy~\cite{numpy}. 
As a reference, we also include the coefficients of a standard Pauli-X encoding with no preprocessing.
The plots show that, on average, different architectures reinforce different frequency spectra.
Standard angle embedding has a spectrum that is peaked and centered in 0 for shallow circuits. It favors low frequencies (with some of them missing) and, increasing the number of layers, the distribution becomes flatter.
Low-frequency encoding further favors lower frequencies over higher ones with respect to the standard Pauli encoding when increasing the number of layers. 
High-frequency encoding, on the other hand, offers a broad frequency spectrum with roughly equal weights on all frequencies. Notably, the 5-layer case presents an exception, with lower frequencies slightly strengthened.
\begin{figure}[h!]
    \centering
    \includegraphics[trim={.cm .cm .cm .cm},clip,
    width=\linewidth]{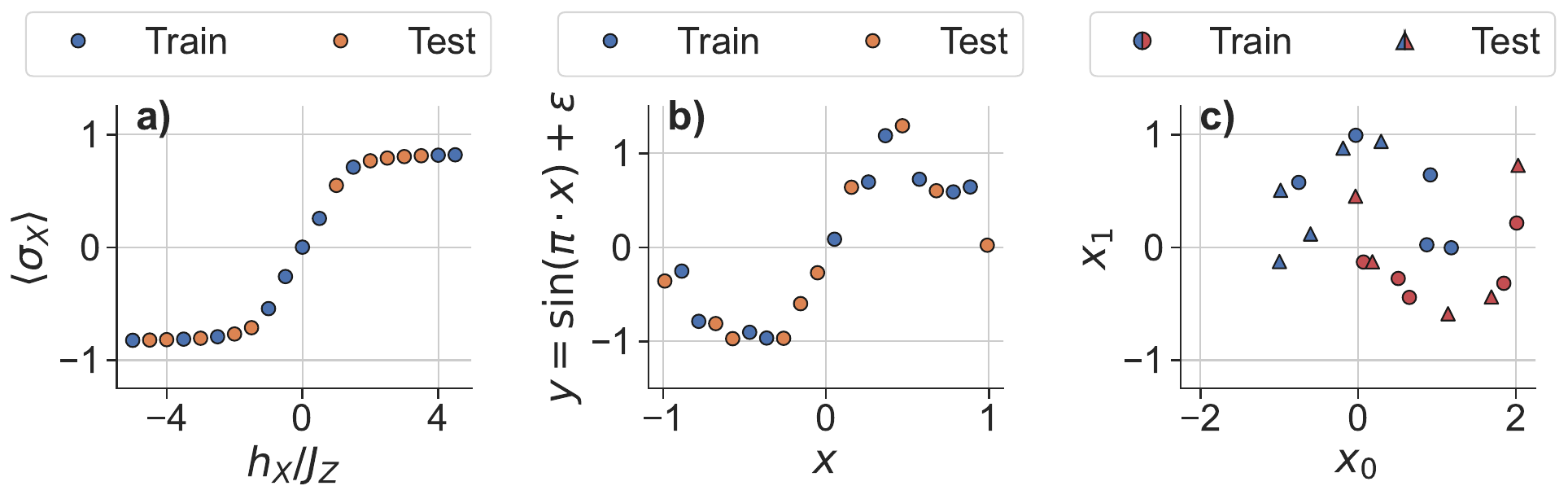}
    \caption{\textbf{Datasets} \textbf{a)} 1D Transverse Field Ising Model (TFIM) dataset. \textbf{b)} Noisy sinusoidal dataset. \textbf{c)} 2-dimensional, nonlinearly separable Moons dataset.}
    \label{fig:datasets}
\end{figure}

\section{Additional numerical results}
\label{appendix: other results}
Here we present further numerical results obtained on the sinusoidal and moons datasets. 

\subsection{Sinusoidal dataset}

\begin{figure}
    \centering
    \includegraphics[width=\linewidth]{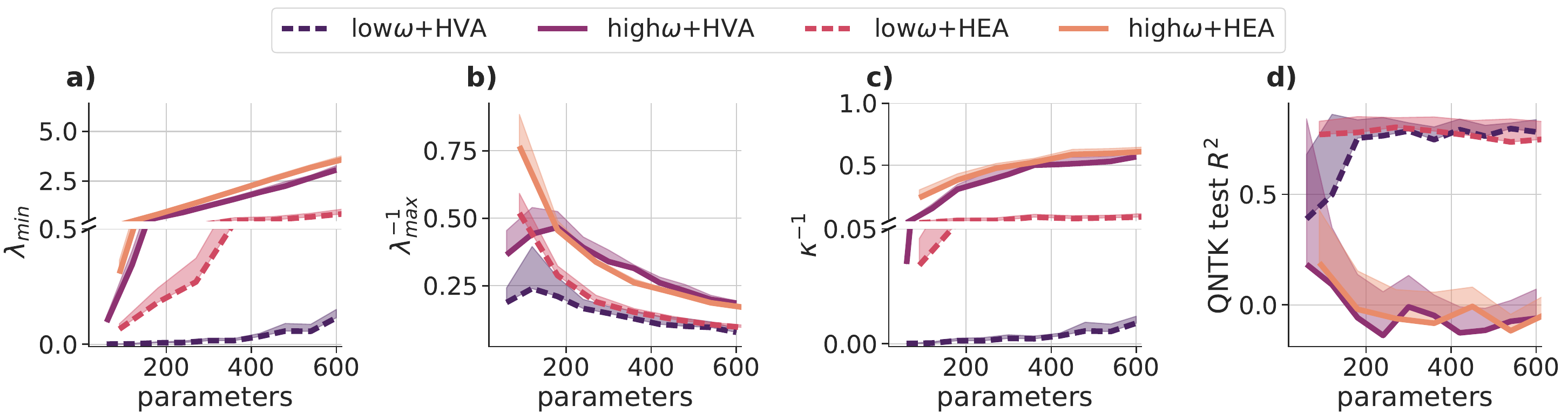}
    \caption{\textbf{QNTK analysis.} This figure summarizes the main features of QNTK for the sinusoidal test case as a function of the number of parameters. The smallest \textbf{a)} and largest \textbf{b)} eigenvalues increase ($1/\lambda_{max}$ decreases), as well as the inverse of the condition number $\kappa^{-1}$ (and hence the typical decay rate $\tau^{-1}=2\kappa^{-1}$) \textbf{c)}. Panel \textbf{d)} represent the $R^2$ score of the fit performed by the QNTK over the whole range of $x$. Lines (solid and dashed) represent the average over 10 different initializations while shaded regions correspond to one standard deviation.}
    \label{fig:sin_NTK_summary}
\end{figure}

The dataset consists of 20 points, distributed over a complete sinusoidal oscillation and subject to Gaussian (white) noise. The analytical function of the function describing the relation under study is 
\begin{equation}
    y = \sin(\pi x) + \epsilon\,,
\end{equation}
where $x \in [-1,1]$ and $\epsilon$ is the stochastic term representing additive white noise with amplitude equal to $0.4$, zero mean and a standard deviation of $0.5$. A representation of the dataset is given in Fig.~\ref{fig:datasets}b. Before being fed to the QNNs, the data labels are rescaled with a min-max scaler (taken from \texttt{sklearn} Python library~\cite{sklearn}) to fit in the range $[-1,1]$. This is required to match the output range of our QNN models, which are only able to provide values in this interval. The inference phase is performed on 100 $x$ points evenly distributed in $[-1,1]$. \\

\begin{figure*}
    \centering
    \includegraphics[
    width=1\textwidth]{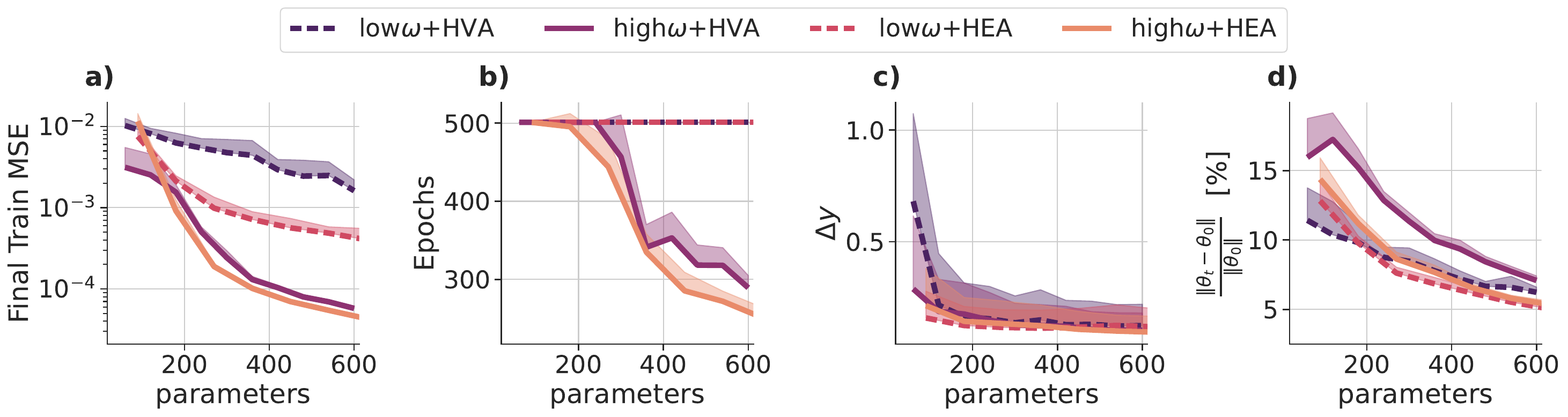}
    \caption{\textbf{Training summary for $\boldsymbol{\eta=\eta_{crit}}$.} Analysis of QNN training on the sinusoidal dataset. \textbf{a)} Final training MSE. \textbf{b)} Training epochs. \textbf{c)} Average absolute difference between QNN and QNTK-based predictions. \textbf{d)} Relative parameter change during training. Lines (solid and dashed) represent the average over 10 different initializations while shaded regions correspond to one standard deviation.}
    \label{fig:sin_summary_max_lr}
\end{figure*}

\begin{figure*}
    \centering
    \includegraphics[width=1.\linewidth]{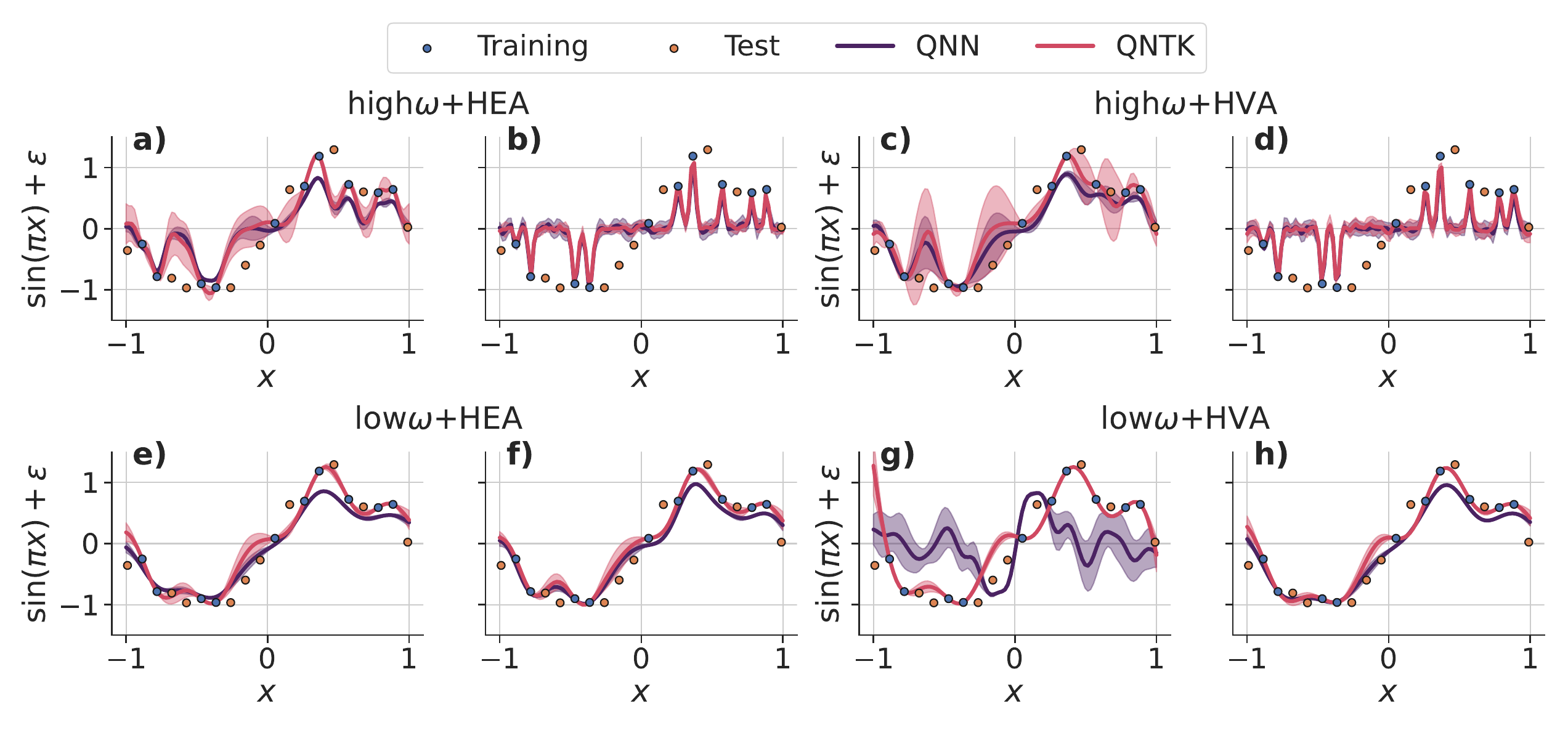}
    
    \caption{\textbf{Inference summary for $\boldsymbol{\eta=\eta_{crit}}$.} For each QNN model design, two panels compare the QNTK inference approximation (red line)  with its trained QNN counterpart (purple line) over the sinusoidal dataset at 5 (left sub-plot) and 50 (right sub-plot) layers. Specifically, we have: high-$\omega$ with HEA using \textbf{a)} 5 and \textbf{b)} 50 layers, high-$\omega$ with HVA using \textbf{c)} 5 and \textbf{d)} 50 layers, low-$\omega$ with HEA using \textbf{e)} 5 and \textbf{f)} 50 layers, low-$\omega$ with HVA using \textbf{g)} 5 and \textbf{h)} 50 layers. Lines represent the average over 10 different initializations while shaded regions correspond to one standard deviation. Blue points denote training data, and orange ones are test samples.}
    \label{fig:5_vs_50_sin}
\end{figure*}

Similarly to the case of the 1D TFIM dataset presented in the main text, this is a regression task for which we employ the same performance metrics. In Fig.~\ref{fig:sin_NTK_summary} we report the outcome of the QNTK diagnostic. For the case of the noisy sinusoidal dataset, the eigenvalue spectrum behaves similarly to what is presented in the main text for the 1D TFIM dataset: eigenvalues grow with the number of parameters (Fig.~\ref{fig:sin_NTK_summary}a and Fig.~\ref{fig:sin_NTK_summary}b) but $\lambda_{min}$ raises faster, determining a descending $\kappa$ and increasing speed of convergence (given by $\kappa^{-1}$) (Fig.~\ref{fig:sin_NTK_summary}c). Moreover, the hierarchy of the QNN models in terms of the convergence speed is respected, with low frequencies models being slower than the others. Furthermore, the diagnosis allows to forecast that the QNNs with high frequencies encoding may overfit training data even for modest depths (see Fig.~\ref{fig:sin_NTK_summary}d). Differently from the 1D TFIM, however, here the low frequencies models do not reach $R^2=1$ for the test set, stopping at approximately 0.75. This could be explained by considering that the presence of Gaussian noise in the dataset %
The QNTK-based diagnosis is also supported by the results presented in Figs.~\ref{fig:sin_summary_max_lr} and \ref{fig:5_vs_50_sin}, where we show the training and inference details. In particular, it is confirmed that high frequencies models hit low final values of training MSE (Fig.~\ref{fig:sin_summary_max_lr}a) while being faster in reaching convergence (Fig.~\ref{fig:sin_summary_max_lr}b). Such low values of the loss function may also suggest that these models are prone to overfitting. 


\subsection{Moons dataset}
\begin{figure}
    \centering
    \includegraphics[width=\linewidth]{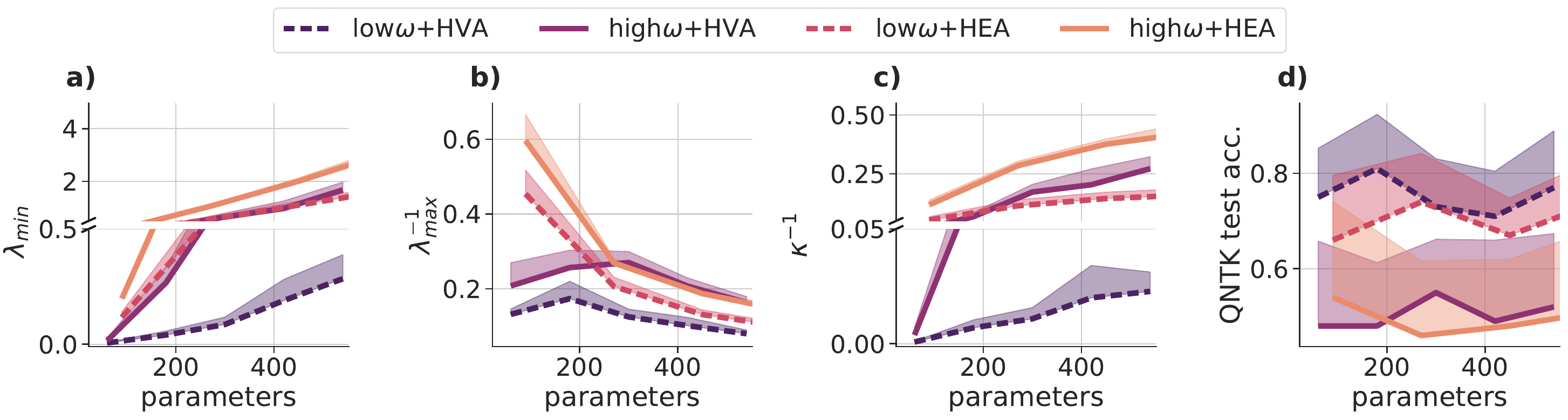}
    \caption{\textbf{QNTK analysis.} This figure summarizes the main features of QNTK for the Moons test case as a function of the number of parameters.  well as the inverse of the condition number $\kappa^{-1}$ (and hence the typical decay rate $\tau^{-1}=2\kappa^{-1}$) \textbf{c)}. Panel \textbf{d)} represent the $R^2$ score of the fit performed by the QNTK over the whole range of features. Lines (solid and dashed) represent the average over 10 different initializations while shaded regions correspond to one standard deviation.}
    \label{fig:moons_NTK_summary}
\end{figure}

\begin{figure*}
    \centering
    \includegraphics[
    width=1\textwidth]{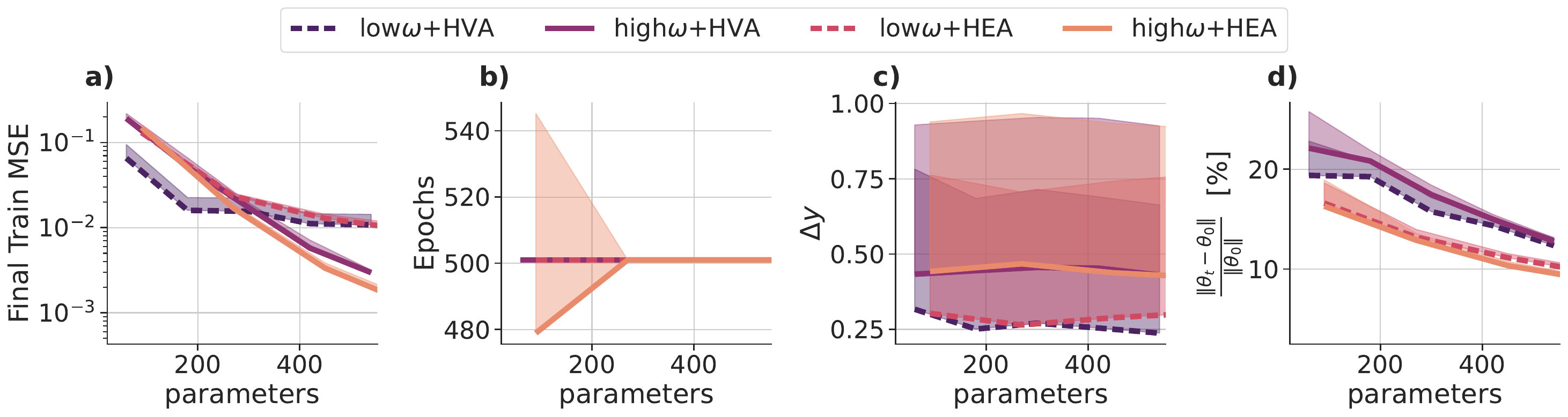}
    \caption{\textbf{Training summary for $\boldsymbol{\eta=\eta_{crit}}$.} Analysis of QNN training on the Moons dataset. \textbf{a)} Final training MSE. \textbf{b)} Training epochs. \textbf{c)} Average absolute difference between QNN and QNTK-based predictions. \textbf{d)} Relative parameter change during training. Lines (solid and dashed) represent the average over 10 different initializations while shaded regions correspond to one standard deviation.}
    \label{fig:moons_summary_max_lr}
\end{figure*}

\begin{figure}
    \centering
    \includegraphics[trim={0.cm 1.cm 1.cm 1.cm},clip,width=\linewidth]{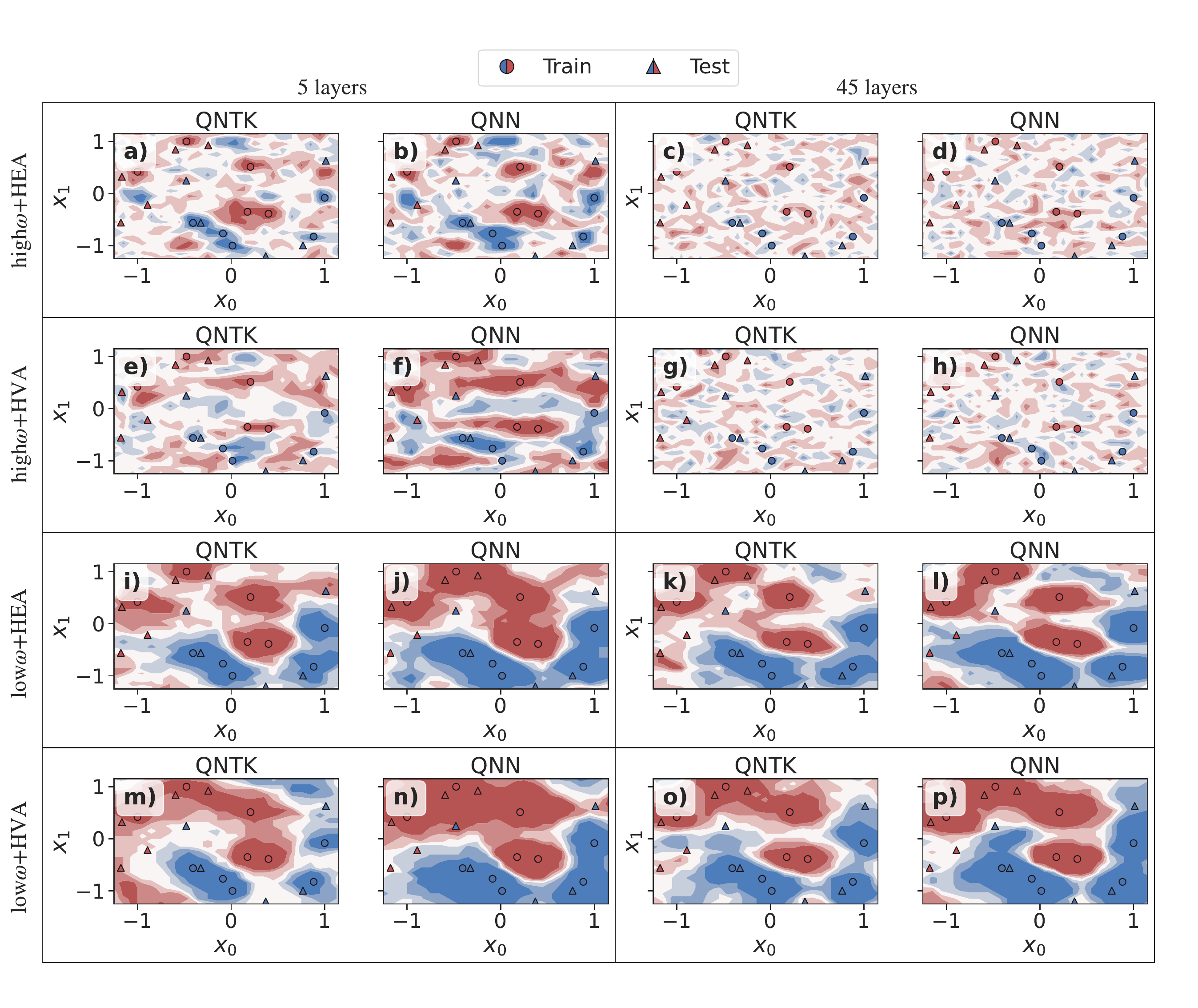}
    \caption{\textbf{Inference summary for $\boldsymbol{\eta=\eta_{crit}}$.} The four columns compare the QNTK inference approximation (first and third columns) with its trained QNN counterpart (second and fourth columns) over the Moons dataset at 5 (first and second columns) and 45 (third and fourth columns) layers for different designs (one per each row). Training and testing data samples are represented by markers, circles and triangles respectively, while the predictions made by the QNN are represented as coloured shades, red and blue depending on the class. The same colouring rule applies to markers. Different nuances of blue and red represent the agreement on the prediction between 10 different training procedures. The stronger the color the higher the agreement.}
    \label{fig:5_vs_50_moons}
\end{figure}

As a last use case, we study a binary classification problem on the Moons dataset. The dataset is 2-dimensional and not linearly separable. 
As for the sinusoidal dataset, we sample $20$ points from the distribution to which we add some Gaussian noise, in this case with amplitude equal to $1$, null mean and a standard deviation equal to $0.2$. The data features initially belong to the intervals $x_0\in[-1,2]$ and $x_0\in[-0.5,1]$. The train-test split is $50\%$, and the dataset is depicted in Fig.~\ref{fig:datasets}c. As this is a classification task, we have one-hot vector labels, and we rescale the data features to fit the interval $[-1,1]$ for training. 
The inference is done on a 2D-grid with steps of $0.1$ spanning the area covered by the dataset for a total of $625$ points.

Due to the increased size of the final inference set, which is more than six times larger than the previous ones, we only study a limited selection of QNN architectures, and specifically thoses with layers $L\in[5, 15, 25, 35, 45]$. 
While for classification tasks a cross-entropy-like loss function is usually employed, in this case we still use MSE to respect the assumption underlying the QNTK derivation. \\

Looking at Fig.~\ref{fig:moons_NTK_summary}a,~b~and~c we can notice a behavior of the QNTK spectrum similar to what has been already observed for previous datasets. The high-frequency models show fast convergence but are expected to perform poorly on unseen data, as shown in Fig.~\ref{fig:moons_NTK_summary}d. This plot also suggests that low-frequency QNNs should generalize better. From Fig.~\ref{fig:moons_summary_max_lr} we can grasp a comprehensive understanding of the training dynamics (plots a~and~b) as well as of the quality of the QNTK-based approximation (plots c~and~d) for $\eta=\eta_{crit}$. For what concerns the training phase, we have two main differences with respect to what we observed so far: the training loss does not reach extremely low values and the number of epochs is constant at the stopping values. Some cases with high frequencies encoding and HEA behave differently, having a total number of epochs lower than 500 but high loss value, suggesting that convergence was achieved because of the large learning rate. This means that the training process should not be considered concluded as there is room for improvement in the training loss minimization. This can also partially explain why the point by point agreement between the QNTK approximation and the QNN predictions over the $625$ points of the grid, reported in Fig.~\ref{fig:moons_summary_max_lr}c, is quite low. %
The other possible reason for such poor performance is that the change in parameters is definitely non-negligible in this case, as we can see in Fig.~\ref{fig:moons_summary_max_lr}d. 

Nevertheless, even if we do not have a precise, quantitative agreement in the predictions, our techniques provide a good qualitative estimate of the QNNs general behavior of the QNNs. 
This can be better understood by looking at Figs.~\ref{fig:5_vs_50_moons}, where we display the average predictions over 10 different initializations made by QNTK and QNN for 5 and 45 layers respectively. 
The predictions do not exactly match and the QNTK seems to overfit training data more than its QNN counterpart. 
This agrees with the fact that we did not conduct training until convergence, and is in fact a particularly instructive example in which the QNTK-based diagnostic would have allowed saving valuable computational resources by suggesting such an early stopping approach. %

\end{document}